\pdfoutput=1
\documentclass{aa}  

\usepackage{graphicx,natbib}
\usepackage{xcolor}
\usepackage{amsmath}
\usepackage{array}
\usepackage{float}
\usepackage[percent]{overpic}
\usepackage{rotating}
\usepackage{gensymb}
\usepackage{tikz}
\usepackage{caption}
\usepackage{placeins}
\usepackage{subcaption}
\usetikzlibrary{positioning}
\def\as {\ifmmode {\rlap.}$\,$''$\,$\! \else ${\rlap.}$\,$''$\,$\!$\fi}
\def\deg {\ifmmode{^{\circ}}\else{$^{\circ}$}\fi}
\usepackage{txfonts}
\begin{document}


   \title{Accretion disk versus jet orientation in H$_{2}$O megamaser galaxies}

 \author{F. Kamali\inst{1}
 	  \fnmsep\thanks{Member of the International Max Planck Research School (IMPRS)
		      for Astronomy and Astrophysics at the Universities of Bonn and Cologne.},
          C. Henkel\inst{1,2}, 
          S. Koyama\inst{3,1},
          C. Y. Kuo\inst{4},
          J. J. Condon\inst{5},
          A. Brunthaler\inst{1},
          M. J. Reid\inst{6},
          J. E. Greene\inst{7},
          K. M. Menten\inst{1},
          C. M. V. Impellizzeri\inst{5,8},
          J. A. Braatz\inst{5}, 
          E. Litzinger\inst{9, 10},
          M. Kadler\inst{9}
          }

   \institute{Max-Planck-Institut f\"{u}r Radioastronomie, Auf dem H\"{u}gel 69, 53121 Bonn, Germany\\              
              \email{fkamali@mpifr-bonn.mpg.de}; 
              \email{fateme.kamali28@gmail.com}
             \and
             Astron. Dept., King Abdulaziz University, P.O. Box 80203, Jeddah 21589, Saudi Arabia\\
             \and
             Academia Sinica Institute of Astronomy and Astrophysics, P.O. Box 23-141, Taipei 10617, Taiwan\\
             \and
             Physics Department, National Sun Yat-Sen University, No. 70, Lien-Hai Rd, Kaosiung City 80424, Taiwan, R.O.C\\
             \and
             National Radio Astronomy Observatory, 520 Edgemont Road, Charlottesville, VA 22903, USA\\
             \and 
             Harvard-Smithsonian Center for Astrophysics, 60 Garden Street, Cambridge, MA 02138, USA\\
             \and
             Department of Astrophysical Sciences, Princeton University, Princeton, NJ 08544, USA \\
             \and
             Joint ALMA Office, Alonso de C{\'o}rdova 3107, Vitacura, Santiago, Chile\\
             \and
             Institut für Theoretische Physik und Astrophysik, Universität Würzburg, Campus Hubland Nord, Emil-Fischer-Str. 31, 97074 Würzburg, Germany\\
             \and 
             Dr. Remeis-Observatory, Erlangen Centre for Astroparticle, Physics, University of Erlangen-Nuremberg, Sternwartstr. 7, 96049 Bamberg, Germany\\
		      }
\titlerunning{Accretion disk versus jet orientation in H$_{2}$O megamaser galaxies}

\authorrunning{Kamali et al.}

   \date{Received August 1, 2015; accepted...}

 
  \abstract
 { An essential part of the paradigm describing active galactic nuclei is the alignment between the radio jet and the associated rotation 
  axis of the sub-pc sized accretion disks. Because of the small linear and angular scales involved, this alignment has not yet been checked in a sufficient number of 
  Low Luminosity Active Galactic Nuclei (LLAGNs).}
{The project intends to examine the validity of this paradigm by measuring the radio continuum on the same physical scale as the accretion disks, to investigate any possible connection
between these disks
   and the radio continuum.}
  {We observed a sample of 18 LLAGNs in the 4.8\,GHz (6\,cm) radio continuum using the Very Long Baseline Array (VLBA) with 3.3 to 6.5 milliarcseconds resolution. 
  The sources were selected to show both
  an edge-on accretion disk revealed by 22\,GHz H$_2$O megamaser emission and signatures of a radio jet. Furthermore, the sources were previously detected in
  33\,GHz radio continuum observations made with the Very Large Array. }
   {Five out of 18 galaxies observed were detected at 8\,$\sigma$ or higher levels (Mrk\,0001, Mrk\,1210, Mrk\,1419, NGC\,2273 and UGC\,3193). While all these sources are known to have maser disks, 
   four of these five sources exhibit a maser disk with known orientation. For all four sources, the radio continuum is misaligned relative to the rotation axis of
   the maser disk, but with a 99.1\% confidence level, the orientations are not random and are confined to a cone within 32\degree of the maser disk's normal. 
   Among the four sources the misalignment of the radio continuum with respect to the normal vector to the maser disk is smaller when the inner radius
   of the maser disk is larger. Furthermore,
   a correlation is observed between the 5\,GHz VLBA radio
   continuum and the [OIII] luminosity and also with the H$_2$O maser disk's inner radius.}
   {}
   
               
		\maketitle
%

\section{Introduction}
Accretion of material onto the central engine of an Active Galactic Nucleus (AGN) is often accompanied by ejection of material, either in the form of outflows or (collimated) jets.
Conservation of angular momentum implies that the distribution of any material ejected from the accretion disk surrounding the supermassive black hole (SMBH) should be 
perpendicular to the disk, unless an external torque is present.
On the other hand, studies have shown that the jet direction and the rotation axis of the large scale galactic disk are not necessarily aligned \citep[e.g.,][]{pringle1999, nagar1999}.
Further studies have been carried out to investigate the alignment of the jet axis with the vector
perpendicular
to the smaller 20-150 pc scale disks
such as dust disks in active galaxies, and have reported that the jets are not always perpendicular
to these disks \citep{schmitt2002}. 
This misalignment indicates that gas on different scales has different angular momentum orientation. 
Whether this is a result of past mergers or warping of the accretion disk, or a general physical condition required for accretion, is still a matter of debate \citep[see][and references therein]{greene2010}. 
With the discovery of H$_2$O maser emission, emitted from the water vapor $J_{K_{-}K_{+}}$=6$_{16}$-5$_{23}$ rotational transition in the sub$\rm -$pc disk surrounding the SMBH in 
NGC\,4258 \citep[e.g.,][]{nakai1993, greenhill1995, miyoshi1995, herrnstein1999}, an opportunity arose to observe the accretion disk of AGNs with the extremely high angular resolution only offered by 
Very Long Baseline Interferometry (VLBI) observations. The other unique 
feature of galaxies with H$_2$O masers in their nuclear accretion disk is that these accretion disks are
viewed within a few degrees edge-on, otherwise linear scales for coherent amplification would be too short, and thus
we expect the putative jets to be on the plane of the sky and free of viewing angle biases.
Therefore, the nuclear region of galaxies hosting maser disks are good laboratories for studying the alignment between the jet and the normal to the pc-size disks.  
\citet{greene2013} and \citet{pjanka2017} investigated this alignment by observing the radio continuum jets on >50 pc scales. 
They also observed the structures such as galaxy-scale disks, bars and spiral arms on <5\,kpc scales and compared the alignment of these structures 
with the smaller scale sub-pc maser disks.
They reported that 
the orientation of the 100 pc scale disks 
and galaxy scale disks are distributed randomly with respect to the orientation of the maser disks.
However, the radio continuum jets (observed on tens of pc to kpc scale) are found to align within $< 15\degree$ of the maser disk's
rotation axis \citep{greene2013}.
With the purpose of investigating the jet-disk alignment on smaller scales ($\sim$ 2\,pc) and also providing a larger sample compared to that already existing in the literature, 
in this work we attempt to study the alignment of the jet and the rotation axis of the accretion disk in
low luminosity active galactic nuclei (LLAGN) with H$_2$O megamaser disks seen approximately edge-on. Recently we have observed a sample of 24 such galaxies with a resolution of $0\as2$ -- $0\as5$
using the Karl G. Jansky Very Large Array (VLA) at 33\,GHz in a pilot project \citep{kamali2017}. Radio synchrotron emission was detected on kpc scales in 21 of these galaxies. In the follow-up
observations presented here, we look
for radio emission on scales of <50 pc ($\sim$5\,mas or $\sim$2\,pc resolution at a fiducial distance of 90 Mpc) using the Very Long Baseline Array (VLBA). We chose C-band
because the synchrotron emission is supposedly brighter at this lower frequency compared to the previous VLA Ka-band observations. We aim at measuring the radio continuum and investigate the
relation between the jet axis and the maser disk orientation.
This paper is organized in the following way:
In Sect.\,\ref{sec:sample} we present our sample. In Sect.\,\ref{sec:data} we present the observations and data reduction process. The description of the observational results is presented in
 Sect.\,\ref{sec:result},
followed by analysis and subsequent discussions in Sect.\,\ref{sec:discussion}. A summary is given in Sect.\,\ref{sec:sum}.

\begin{center}
\begin{figure*}[h!]
\begin{tikzpicture}
\node (img1)  {\includegraphics[width=.47\textwidth]{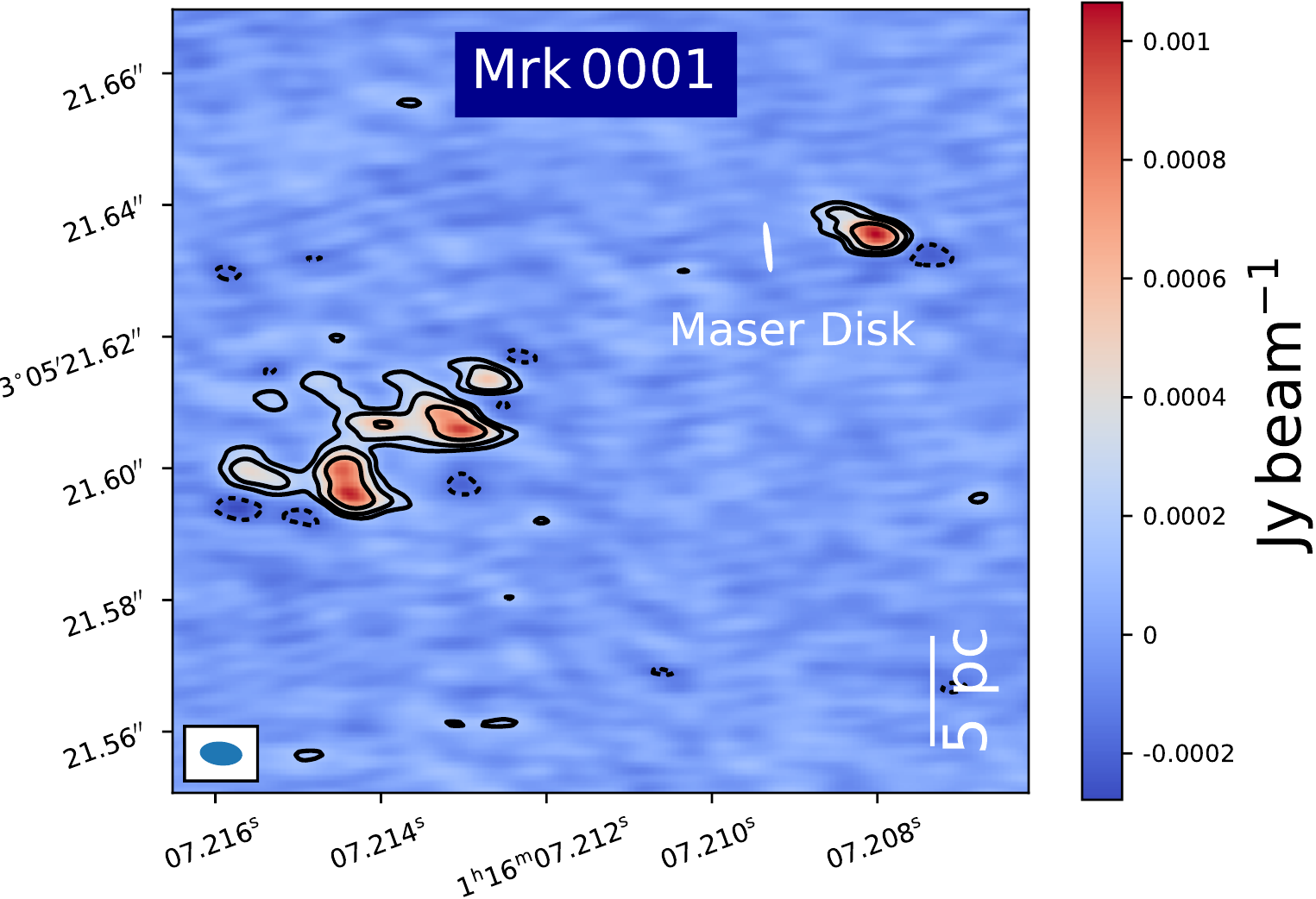} \hspace{0.2cm} \includegraphics[width=.47\textwidth] {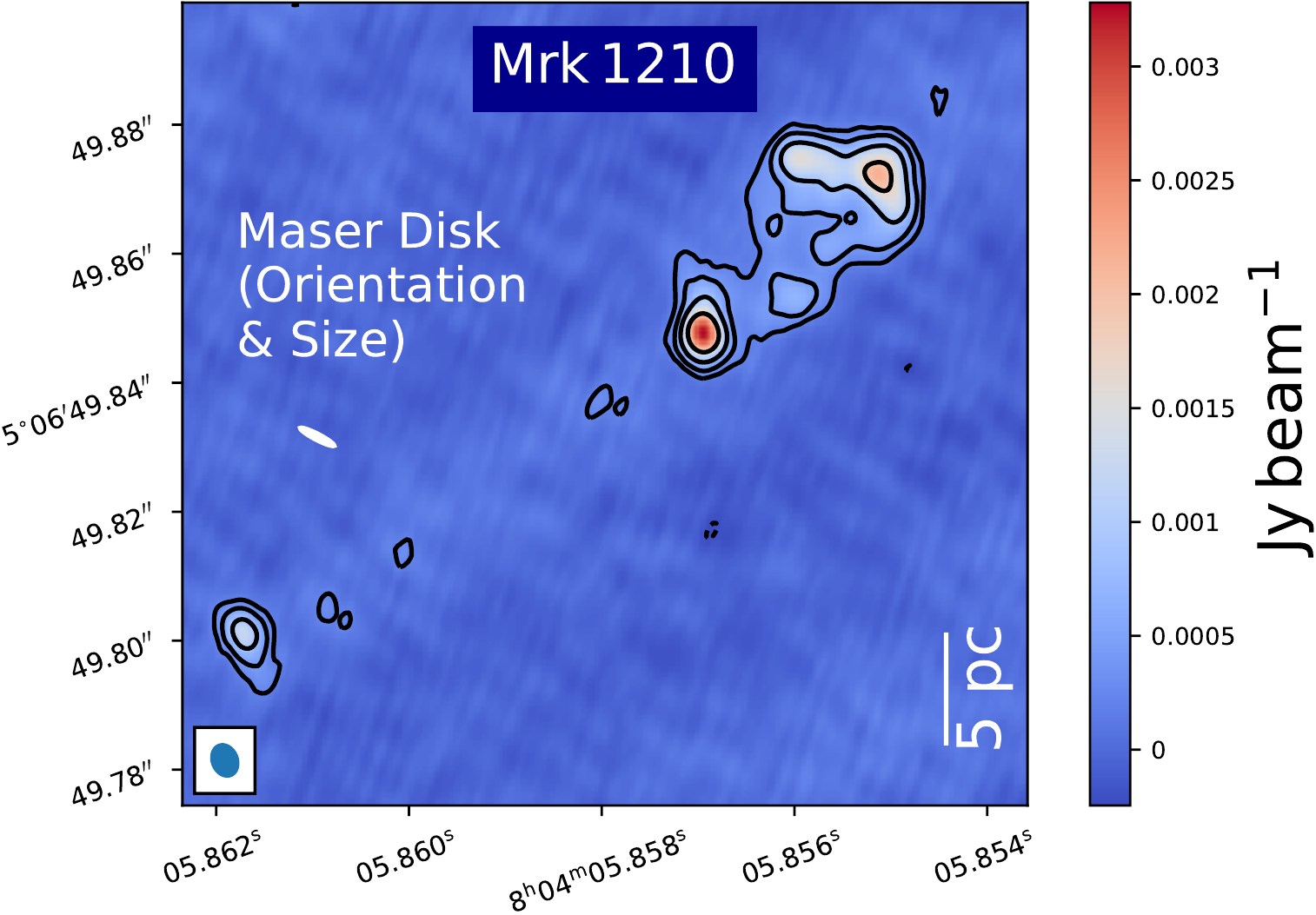}}; 
\node [node distance=0.1cm, below of= img1,yshift=-6.2cm] (img2)  {\includegraphics [width=.47\textwidth]{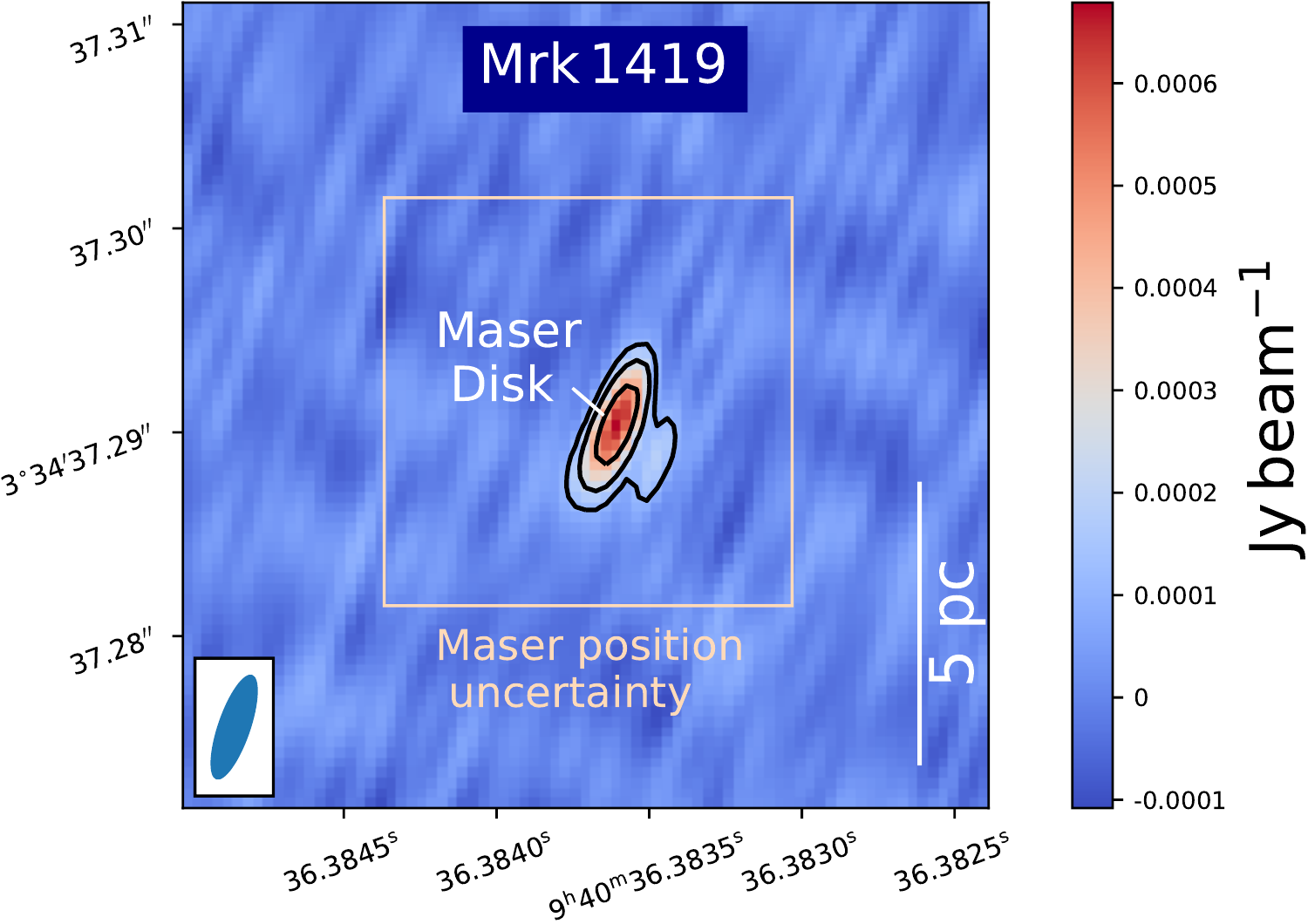}\hspace{0.2cm}\includegraphics[width=.47\textwidth]{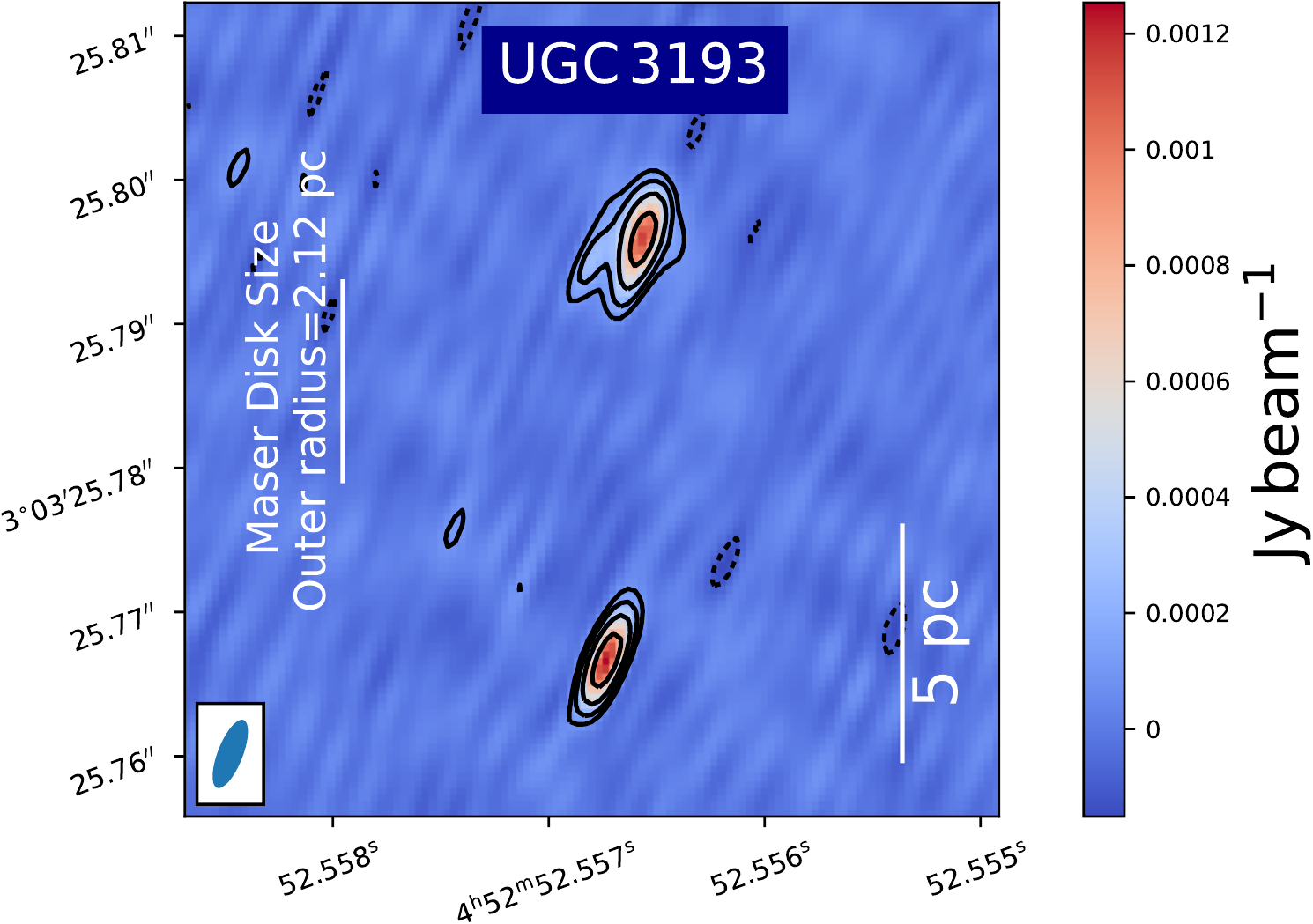}};
 \node [node distance=0.1cm, below of= img2,yshift=-6.2cm](img3) {\includegraphics[width=.47\textwidth]{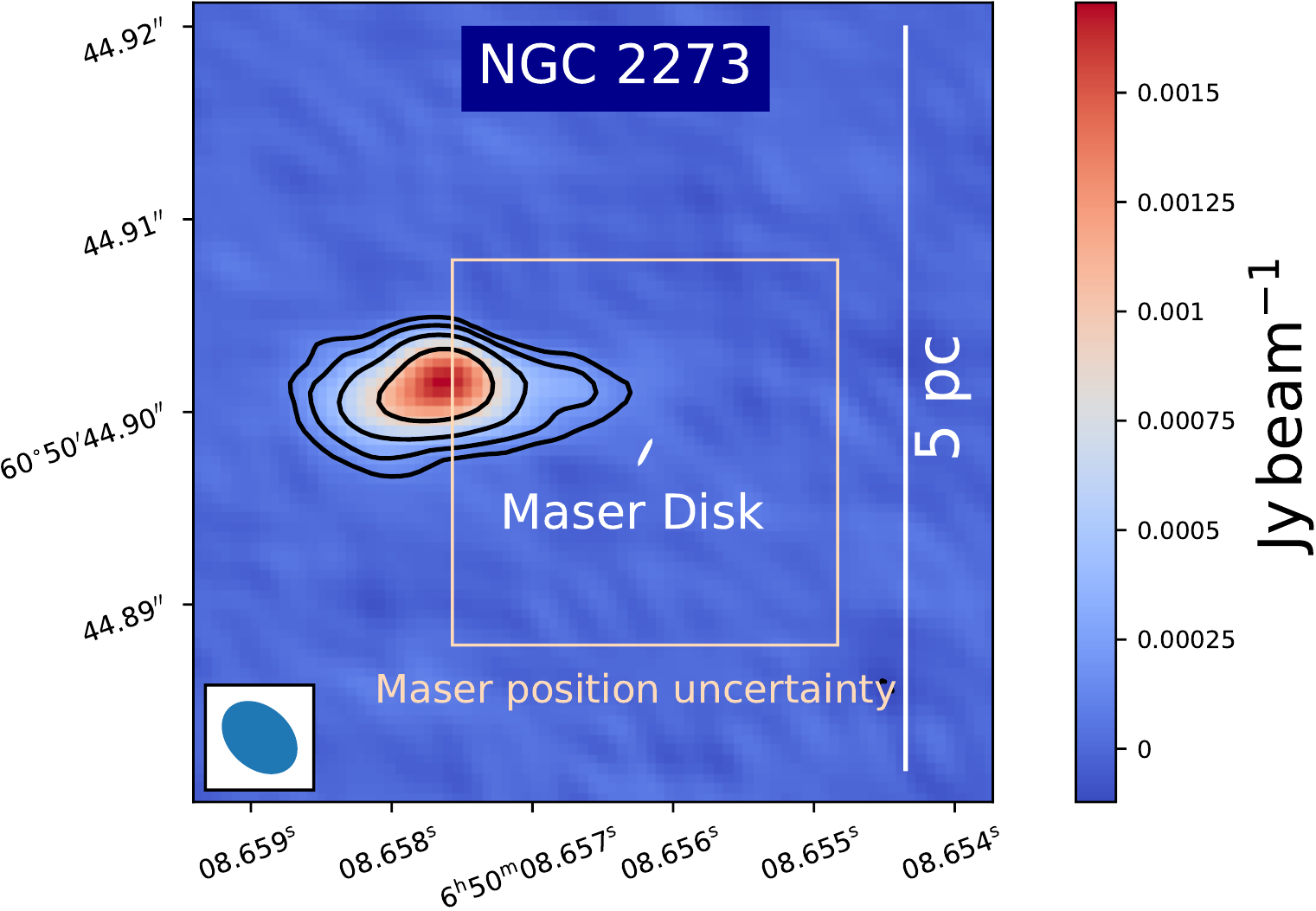} \hspace{0.2cm} \includegraphics[width=.48\textwidth]{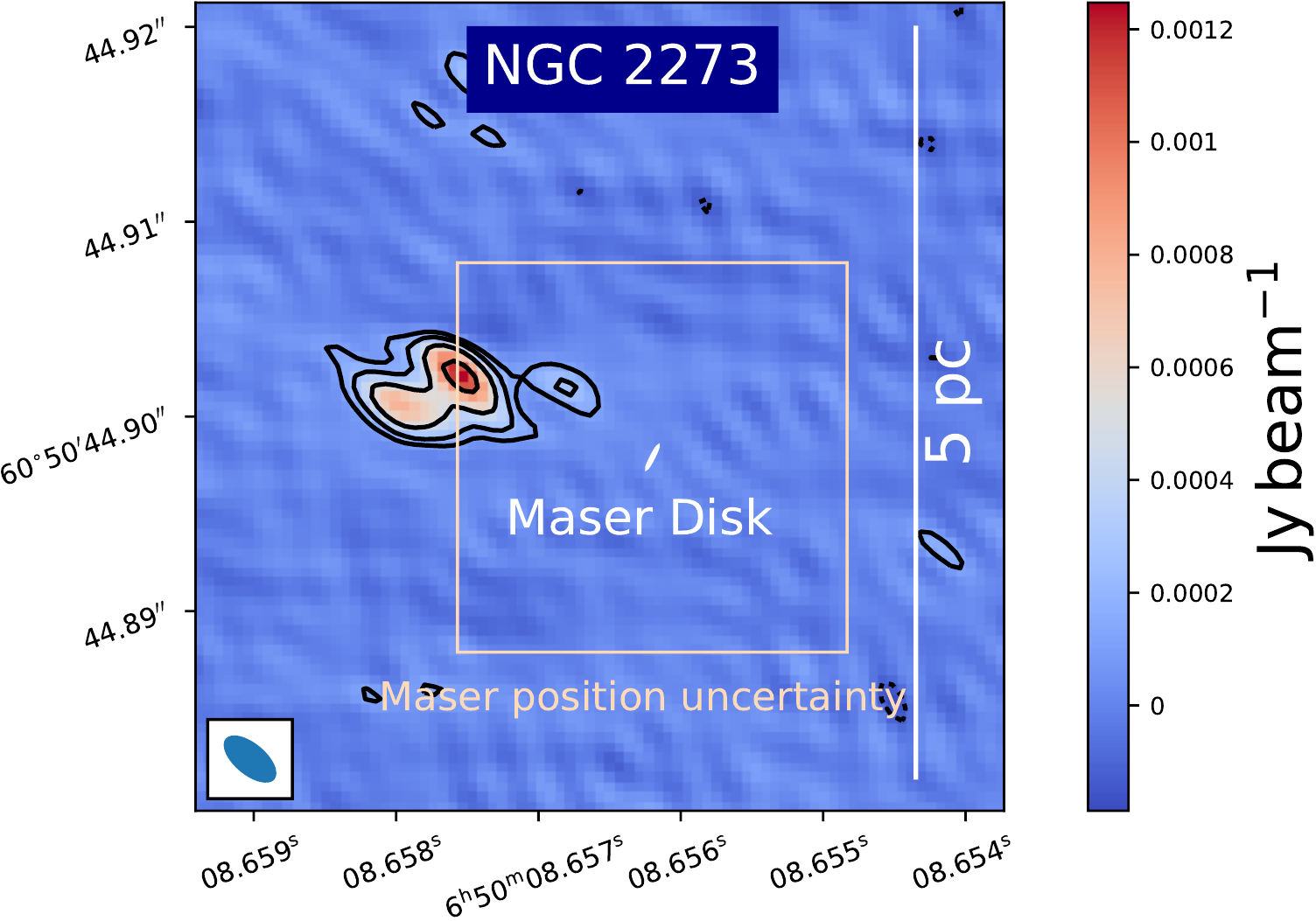}};
 \node[left=of img3, node distance=0cm, rotate=90, anchor=center, xshift=5cm,font=\color{black}\LARGE] {Declination\,};
 \node[below=of img3, node distance=0cm, yshift=1cm, font=\color{black}\LARGE] { Right Ascension\,};
 \end{tikzpicture}
 \caption[Contour Maps]{The 5\,GHz contour maps. The contour levels are $\pm$3, 6, 12 and 24 times the rms given in Table\,\ref{table:properties}.
 The synthesized beam is shown in the lower left corner of each plot. For UGC\,3193 the maser disk size, for Mrk\,1210 the maser disk size and orientation and
 for Mrk\,0001, Mrk\,1419 and NGC\,2273 the maser disk size, orientation, position and position uncertainties (light orange squares) are shown. For Mrk\,0001 the maser position
 uncertainty is small. For Mrk\,1210 and UGC\,3193 the positions of the maser disks are not yet known.
 For NGC\,2273, two contour maps using different Briggs weighting functions are shown: robust parameter 5 which is a pure natural weighting (left)
 and 0 which is between natural and uniform (right).}\label{fig:images1}
\end{figure*}
\end{center}


\section{The sample}\label{sec:sample}
We selected a sample of 18 LLAGNs with declinations > -5\,\degree, including 14 sources from our 33\,GHz survey \citep[][]{kamali2017} plus four more from the
Megamaser Cosmology Project \citep[MCP, e.g.,][]{reid2009, reid2013, braatz2010, greene2010, kuo2011, kuo2013, 
pesce2015, gao2017}.
These sources have signatures of megamasers disks, i.e., they show three groups of H$_2$O megamaser emissions in their spectrum; one set of maser features have velocities grouped around the recession
velocity of the galaxy and those of the two other sets of maser features are offset from this recession velocity, representing the approaching and receding sides of the disk.
In our selected sample, 12 sources are considered ``clean disks'', where the maser emission from the disk
dominates over any other maser emission from the jets or outflows \citep{pesce2015}. These sources are shown in boldface in all tables.
The maser emission in six other sources of our sample may not arise from the disk, but rather from
the nuclear outflows or
star forming regions in these galaxies. Three out of these six non-clean disks are detected in this work. The maser distributions in two of these three galaxies (Mrk\,0001 and Mrk\,1210) have been mapped using the VLBI and the consequent fitting of disk
models does not rule out a disk maser scenario (we refer the reader to Sect.\,\ref{sect:individual sources} for further detail on the maser disk observations for the individual sources). Therefore, since it is unclear whether 
either of the aforementioned scenarios applies (the maser emission may actually arise from the disks), we
consider all the sources as maser disks and our analysis is based on this assumption. The MCP is an ongoing project and the maser disk properties such as disk sizes and orientations
are not yet measured for all the candidates.

For instance, the distances measured by the MCP (which are among the most accurate ones), are only available for four sources in our sample. These distances are in agreement within uncertainties
with the distances listed in the NASA$/$IPAC Extragalactic Database (NED)\footnote{\url{https://ned.ipac.caltech.edu/}}. For consistency we adopted the NED distances 
since they are available for all sources in our sample.
The NED 
distances were obtained using H$_0$=70.0\,km\,s$^{-1}$\,Mpc$^{-1}$, $\Omega_{matter}$=0.27 and $\Omega_{vacuum}$=0.73 as the cosmological parameters. 
H$_2$O maser luminosities from the literature were also rescaled to H$_0$=70.0\,km\,s$^{-1}$\,Mpc$^{-1}$ to be consistent with other luminosities in this work.

  \begin{table*}[ht]
\caption {The VLBA observations.} \label{table:calibrators}
{\tiny
\begin{center}

\begin{tabular}{l  l c c c c   c} 
\hline \hline
& & & & & \tabularnewline
\textbf{Galaxy}  & \textbf{Right Ascension/ Declination} & \textbf{Fringe Finder} & \textbf{Calibrator} 
& \textbf{RA/Dec}  \tabularnewline
& &  &   & \textbf{uncertainty}  \tabularnewline
&    \textbf{$J$2000}               &          &  &  (arcsec)       \tabularnewline
\hline
& & & &  \tabularnewline
 Mrk\,0001/NGC\,0449    & $01^{\rm h}16^{\rm m}07^{\rm s}.20/+33\deg05'21\as75$ & 2007+777/ 3C84  & J0137+3122/J0112+3208 &         0.02  /  0.02      \tabularnewline
 \textbf{J0126-041}7             & $01^{\rm h}26^{\rm m}01^{\rm s}.64/-04\deg17'56\as23$ & 2007+777/ 3C84  & J0123-0348 &                    0.04 /  0.04       \tabularnewline 
 \textbf{NGC\,0591/Mrk\,1157}    & $01^{\rm h}33^{\rm m}31^{\rm s}.23/+35\deg40'05\as79$ & 2007+777/ 3C84  & J0137+3309 &                    0.10 /  0.11       \tabularnewline
%
 Mrk\,1029              & $02^{\rm h}17^{\rm m}03^{\rm s}.57/    +05\deg17'31\as15$ & 2007+777/ DA193  & J0215+0524 &                   0.18 /  0.18       \tabularnewline
  \textbf{NGC\,1194}              & $03^{\rm h}03^{\rm m}49^{\rm s}.11/   -01\deg06'13\as48$ & 2007+777/ DA193  & J0259-0019 &                   0.03  /  0.03      \tabularnewline
 J0350-0127             & $03^{\rm h}50^{\rm m}00^{\rm s}.35/   -01\deg27'57\as40$ & 2007+777/ DA193  & J0351-0301 &                   0.15  / 0.15       \tabularnewline
 J0437+6637             & $04^{\rm h}37^{\rm m}08^{\rm s}.28/  +66\deg37'42\as10$ & 2007+777/ DA193 & J0449+6332/ J0429+6710 &        0.12  /  0.12      \tabularnewline
 UGC\,3193              & $04^{\rm h}52^{\rm m}52^{\rm s}.56/    +03\deg03'25\as52$ & 2007+777/ DA193 & J0453+0128 &               0.11 / 0.29        \tabularnewline
  \textbf{NGC\,2273}              & $06^{\rm h}50^{\rm m}08^{\rm s}.69/    +60\deg50'45\as10$ & 2007+777/ DA193 & J0638+5933 &                    0.28 / 0.27        \tabularnewline
 \textbf{UGC\,3789}              & $07^{\rm h}19^{\rm m}30^{\rm s}.95/    +59\deg21'18\as37$ & 3C84/ 4C39.25 & J0737+5941 &                      0.08 / 0.08        \tabularnewline
 \textbf{Mrk\,0078}              & $07^{\rm h}42^{\rm m}41^{\rm s}.73/    +65\deg10'37\as39$ & 3C84/ 4C39.25 & J0737+6430 &                      0.03 / 0.03        \tabularnewline
 
 Mrk\,1210/Phoenix      & $08^{\rm h}04^{\rm m}05^{\rm s}.86/    +05\deg06'49\as83$ & DA193/ 4C39.25/ 3C345 & J0803+0421 &              0.03  / 0.03       \tabularnewline
 
 \textbf{Mrk\,1419/NGC\,2960}    & $09^{\rm h}40^{\rm m}36^{\rm s}.38/    +03\deg34'37\as36$ & 3C84/ 4C39.25 & J0930+0034 &                      0.14   / 0.13        \tabularnewline
 \textbf{NGC\,4388}              & $12^{\rm h}25^{\rm m}46^{\rm s}.78/    +12\deg39'43\as77$ & DA193/ 4C39.25/ 3C345 & J1225+1253 &              0.02   / 0.02      \tabularnewline
 \textbf{NGC5765b}	         & $14^{\rm h}50^{\rm m}51^{\rm s}.5/     +05\deg06'52\as0$ & DA193/ 4C39.25/ 3C345 & J1458+0416 &              1.25/ 1.25         \tabularnewline
 \textbf{UGC\,9639/Mrk\,0834} 	& $14^{\rm h}58^{\rm m}35^{\rm s}.99/  +44\deg53'01\as0$  & 4C39.25/ 2007+777  & J1459+4442 &                   0.5/ 0.5           \tabularnewline  
 \textbf{NGC\,6264}		& $16^{\rm h}57^{\rm m}16^{\rm s}.13/  +27\deg50'58\as5$  & 4C39.25/ 2007+777  & J1659+2629 &                   0.5/ 0.5           \tabularnewline
 \textbf{NGC\,6323}		& $17^{\rm h}13^{\rm m}18^{\rm s}.07/  +43\deg46'56\as8$  & 4C39.25/ 2007+777  & J1707+4536 &                   0.5/ 0.5           \tabularnewline
 \hline   
 \hline
  \\
 Mrk\,0001 		& $01^{\rm h}16^{\rm m}07^{\rm s}.209/ +33\deg05'21\as634$  &  3C84              & J0119+3210		& 0.001/ 0.001	\tabularnewline 
 UGC\,3193		& $04^{\rm h}52^{\rm m}52^{\rm s}.5569/ +03\deg03'25\as768$  &  3C84              & J0459+0229		& 0.001/ 0.001	\tabularnewline
 Mrk\,1210		& $08^{\rm h}04^{\rm m}05^{\rm s}.8570/ +05\deg06'49\as846$   & 4C39.25/ 3C84      & J0803+0421		& 0.0008/ 0.002	\tabularnewline
 \textbf{Mrk\,1419}		& $09^{\rm h}40^{\rm m}36^{\rm s}.3835/ +03\deg34'37\as289$  & 4C39.25/ 3C84      & J0930+0034		& 0.001/ 0.001	\tabularnewline

 \hline
\end{tabular}\par
 
\end{center}

\bigskip
\textbf{Notes}. Column 1: name of galaxy. The sources in bold hold clean maser disks (see Sect.\,\ref{sec:sample}). Column 2: $J$2000 right ascension and declination. For NGC\,5765b, NGC\,6323, NGC\,6264 and UGC\,9639
the coordinates are taken from NED and for the remaining sources the coordinates are from \citet{kamali2017}. Column 3: fringe finders. Column 4: phase calibrators. 
Column 5: right ascension and declination uncertainties.
The second VLBA observations are presented in the lower part. The pointing coordinates for the second observations were taken from the first run of VLBA observations. 

}   
\end{table*}
 
\section{Observations and data reduction}\label{sec:data}
Our sample was observed between October and November 2015 using the VLBA in C-band. The synthesized beamwidth ranges from 3.3 to 6.5 mas. We used a total bandwidth of 128\,MHz
(2$\times$64\,MHz IFs, 4.644 to 4.708\,GHz and 4.836 to 4.900\,GHz) and right-hand circular polarization.
The sources were grouped into 6 sets of three for observation, with a total time of 7 hours per group
and an integration time of $\sim$1.5 hours per target after (i.e., excluding) calibration. Our observations included two scans of fringe finders, one placed at the beginning and one at the end of the observations.
The sources and their associated calibrators were observed alternating in scans of $\sim$1 minute. 
Table\,\ref{table:calibrators} indicates fringe finders and calibrators, and the $J$2000 coordinates obtained from previous VLA observations.
From the 18 observed sources, five were detected at signal-to-noise ratios (S/Ns) of 8 or higher levels. In 
the case of four of these detected sources, Mrk\,0001, Mrk\,1210, Mrk\,1419 and UGC\,3193,
the initial images did not have a sufficiently high quality. Therefore, to reassure our detections and obtain higher S/Ns, we re-observed them with the VLBA in C-band, between October and November 2017. 
This time they were grouped in 2 sets of 2 sources,
with a total time of 8 hours per group (2.5 hours on source per target, after calibration), bandwidth of 256\,MHz (8$\times$32\,MHz IFs, from 4.836 to 5.092\,GHz) and dual polarization.

The data were calibrated using the Astronomical Image Processing System (AIPS) software package developed by the National Radio Astronomy
Observatory (NRAO). 
At first, the ionospheric dispersive delays and the Earth orientation parameters were corrected using AIPS tasks VLBATECR and VLBAEOPS respectively. Then, using task CLCOR, the parallactic angle
correction was applied. The visibilities were normalized by the auto-correlations with the task ACCOR. The instrumental phases and delays were removed using the fringe finders listed in Table\,\ref{table:calibrators}
and task VLBAMPCL. Amplitude corrections were applied using system temperature and the antenna gain information by the APCAL task in AIPS. 
We solved phases, delays and rates by performing fringe fitting on the calibrators listed in Table\,\ref{table:calibrators} assuming a point source model, and applied the solutions to the
targets. 

For imaging, we used the task IMAGR in AIPS. Different weighting functions (including
tapering of the uv-distances) were tried to find the optimum solution for a better S/N. 

\newcolumntype{H}{>{\setbox0=\hbox\bgroup}c<{\egroup}@{}}
\begin{table*}[t]
\caption [The sample]{Obtained properties for detected sources. } \label{table:obs_properties}
{\tiny
\begin{center}
\begin{tabular}{l c c c l l l l l l l l c} 
\hline \hline
& & & & & & & & &\tabularnewline
\textbf{Galaxy}   &  \textbf{5\,GHz peak flux} & \textbf{5\,GHz integrated flux}& \textbf{Right Ascension/ Declination} &\textbf{Uncertainties} &\textbf{$\alpha^{33}_{5}$} & $\rm T_b$ & \textbf{Putative}\tabularnewline
	              & (mJy/beam)&    (mJy)  & \textbf{}& (mas/mas) & & (K) & \textbf{core}\tabularnewline
&  first obs || second obs&first obs || second obs  & & & &  & \tabularnewline
\hline
& & & & & & \tabularnewline

Mrk\,0001             	 &    1.30$\pm$0.09 || 1.05 $\pm$ 0.05     &   8.6$\pm$0.8 || 8.9$\pm$0.3     & $01^{\rm h}16^{\rm m}07^{\rm s}.208/ +33\deg05'21\as636$    & 0.018/ 0.017   & 0.39$\pm$0.03 & $6.5\times10^6$ & NW  \tabularnewline
Mrk\,1210             	 &    0.92$\pm$0.07 || 3.10 $\pm$ 0.07     & 15.05$\pm$0.8 || 17.9$\pm$ 0.7   & $08^{\rm h}04^{\rm m}05^{\rm s}.8570/ +05\deg06'49\as848$   & 0.0008/ 0.002  & 0.40$\pm$0.01 & $1.3\times10^7$ & CN\tabularnewline 
\textbf{Mrk\,1419}       &    0.94$\pm$0.06 || 0.66 $\pm$ 0.03     &   1.7$\pm$0.2 || 0.7$\pm$0.3     & $09^{\rm h}40^{\rm m}36^{\rm s}.3835/+03\deg34'37\as289$    &   0.001/ 0.001 & 0.31$\pm$0.09 & $>2.6\times10^6$&...\tabularnewline

\textbf{NGC\,2273}       &    1.6 $\pm$ 0.03 || ...                & 3.3$\pm$0.1  || ...              & $06^{\rm h}50^{\rm m}08^{\rm s}.6577/  +60\deg50'44\as9012$ &  0.0005/ 0.0006 & 0.11$\pm$0.06 & $9.1\times10^6$ & ... \tabularnewline

UGC\,3193                &    0.7$\pm$0.06 || 1.3 $\pm$ 0.03       & 2.0$\pm$0.1  ||  3.0$\pm$0.9     & $09^{\rm h}40^{\rm m}36^{\rm s}.3836/ +03\deg34'37\as290$   & 0.0004/ 0.001   & -0.14$\pm$0.04 & $1.3\times10^7$ &N \tabularnewline
\hline
                                                                                                                                                                                                                                                 
\end{tabular}\par
\end{center}
\bigskip
\textbf{Notes}. Column 1: name of galaxy. The sources in bold hold clean maser disks (see Sect.\,\ref{sec:sample}).
Column 2: 5\,GHz peak flux density of the brightest component.
Column 3: 5\,GHz integrated flux. For detected sources with multiple components the reported flux is the sum of the fluxes from all the components.
Column 4: $J$2000 right ascension and declination obtained (except for NGC\,2273) from the second observations for the peak position of the brightest component.
Column 5: right ascension and declination uncertainties. 
Column 6: Spectral index obtained from 33\,GHz VLA observation and 5\,GHz VLBA observations assuming a power law dependence of S $\propto$ $\rm \nu^{-\alpha}$.
Column 7: Brightness temperatures. If the source is unresolved, a lower limit is given.
Column 8: Putative core of the AGN, corresponding to the component with highest brightness temperature. NW stands for northwestern, CN for central and N for northern.

}                                                                       
\end{table*}


\newcolumntype{H}{>{\setbox0=\hbox\bgroup}c<{\egroup}@{}}
\begin{table*}[t]
\caption [The sample]{Sample properties. } \label{table:properties}
{\tiny
\begin{center}
\begin{tabular}{l c c c c l l l l l l} 
\hline \hline
& & & & & & & & &\tabularnewline
\textbf{Galaxy}   &  \textbf{Distance} & \textbf{5\,GHz upperlimits}& \textbf{rms} &\textbf{33\,GHz flux}
 & \textbf{Hard X-ray flux}&\textbf{log L$_{H_2O}$}  &\textbf{Type of}\tabularnewline
	              & (Mpc)&    (mJy)  & ($\mu$Jy/beam)& (mJy) & ($\mathrm{10^{-11}erg\,s^{-1}\,cm^{-2}}$) & (L$_\odot$)&\textbf{activity}\tabularnewline
& & &first obs || second obs  & & & & \tabularnewline
\hline
& & & & & & \tabularnewline
\textbf{J0126-0417}            &     76.2 $\pm$ 5.4     & < 0.13                             & 42.5          & 0.14 $\pm$ 0.02      & <0.34 			& 2.1   & $U$	 \tabularnewline 
J0350-0127            &    174.2 $\pm$ 12.2    & < 0.13                             & 37.6          & 0.17 $\pm$ 0.04      & 0.45$^{+0.17}_{-0.16}$ 	& 3.7   & $U$ 	\tabularnewline
J0437+6637            &     52.9 $\pm$ 3.7     & < 0.13                             & 36.8          & 0.11 $\pm$ 0.02      & 0.29$\pm$0.14 		& 1.4   & $U$ 	\tabularnewline

Mrk\,0001             &     64.2 $\pm$ 4.5     &   ...      & 45.6 || 49.7    & 3.97 $\pm$ 0.13      & <0.32 			& 1.9   & Sy2	\tabularnewline
\textbf{Mrk\,0078}             &    159.9 $\pm$ 11.2    & < 0.10                             & 33.4          & 1.69 $\pm$ 0.11      & 0.63 $\pm$ 0.18		& 1.6   & Sy2 	\tabularnewline
Mrk\,1029             &    126.0 $\pm$ 8.8     & < 0.11                             & 36.7          & 1.29 $\pm$ 0.16      & <0.35		        & 2.8   & $U$ 	\tabularnewline
Mrk\,1210             &     61.3 $\pm$ 4.3     & ...     & 80.0 || 32.0   & 8.14 $\pm$ 0.43      & 3.58$\pm$0.20		& 2.0   & Sy2, Sy1 \tabularnewline
\textbf{Mrk\,1419}            &     75.3 $\pm$ 5.3     &...        & 39.0 || 37.0    & 0.36 $\pm$ 0.07      & <0.34		        & 2.7   & LINER \tabularnewline

\textbf{NGC\,0591}             &     61.1 $\pm$ 4.3     & < 0.11                             & 41.3          &  1.52 $\pm$ 0.12     & 0.37$^{+0.12}_{-0.11}$ 	& 1.5   & Sy2	\tabularnewline
\textbf{NGC\,1194}             &     55.4 $\pm$ 3.9     & < 0.12                             & 38.5          &  1.05 $\pm$ 0.04     & 2.21$\pm$0.18		& 2.8   & Sy\,1.9\tabularnewline
\textbf{NGC\,2273}             &     26.8 $\pm$ 1.9     & ...                        & 39.7          &  3.04 $\pm$ 0.36     & 0.67 $\pm$ 0.16 	& 0.9   & Sy2	\tabularnewline
\textbf{NGC\,4388}             &     40.8 $\pm$ 2.9     & < 0.13                             & 44.6          &  3.01 $\pm$ 0.08     & 15.81$^{+0.16}_{-0.15}$	& 1.2   & Sy2 	\tabularnewline
\textbf{NGC\,5765b}	      &     122.0 $\pm$ 8.5    & < 0.11                             & 37.2          & ...		      &	0.40$^{+0.17}_{-0.18}$		& 3.6	& $U$     \tabularnewline
\textbf{NGC\,6264}	      &     145.4 $\pm$ 10.2   & < 0.12                             & 38.5          & ...		      &	<0.32			& 3.1	& Sy2     \tabularnewline
\textbf{NGC\,6323}             &     110.6 $\pm$ 7.8    & < 0.12                             & 39.1          & ...		      &	<0.30			& 2.7	& Sy2     \tabularnewline

UGC\,3193             &     63.2 $\pm$ 4.4     & ...     & 31.3 || 34.0    & 0.80 $\pm$ 0.05      & <0.39 			& 2.5   & $U$ 	\tabularnewline
\textbf{UGC\,3789}             &     47.4 $\pm$ 3.3     & < 0.11                             & 35.0          & 0.20 $\pm$ 0.02      & 0.26$^{+0.14}_{-0.13}$	& 2.7   & Sy2   \tabularnewline
\textbf{UGC\,9639}             &     155.9 $\pm$ 10.9   & < 0.09                             & 30.0          &  ...		      &	<0.26	                & 2.6	& LINER      \tabularnewline
\hline
                                                                                                                                                                                                                                                 
\end{tabular}\par
\end{center}
\bigskip
\textbf{Notes}. Column 1: name of galaxy. The sources in bold hold clean maser disks (see Sect.\,\ref{sec:sample}).
Column 2: Hubble flow distances (relative to the 3\,K Cosmic Microwave Background) in Mpc, assuming H$_0$=70\,km\,s$^{-1}$\,Mpc$^{-1}$ (see Sec.\,2).
Column 3: A 3$\sigma$ upper limit for the nondetections in the VLBA 5\,GHz observations.
Column 4: rms of the 5\,GHz maps.
Column 5: 33\,GHz integrated flux densities with uncertainties \citep{kamali2017}. 
Column 6: Swift/BAT hard X-ray fluxes (20-100\,keV) with uncertainties (Litzinger\,et\,al.in\, prep.). 
Column 7: logarithm of water maser luminosity in units of solar luminosity \citep[][modified for H$_0$=70\,km\,s$^{-1}$\,Mpc$^{-1}$]{zhang2012}. The NGC\,5765b and UGC\,9639
data are from \citet{kuo2018}.
Column 8: types of nuclear activity following NED; $U$ stands for unidentified.
}                                                                       
\end{table*}


 \begin{table*}[t]
\caption [Position angles (PAs).]{Position angles (PAs). } \label{table:PAs}
{\tiny
\begin{center}
\begin{tabular}{l c c c c c c l } 
\hline \hline
& & & & & & \tabularnewline
\textbf{Galaxy}  & & \textbf{maser disk}& \textbf{5\,GHz jet direction}& \textbf{jet offset} & \textbf{33\,GHz continuum} &\textbf{galaxy scale} & \textbf{ref}\tabularnewline
	         & &\centering{(degree)} & (degree) &   (degree)  & (degree) & (degree)\tabularnewline
\hline
& & & & &  \tabularnewline
Mrk\,0001 &&  6.4$\pm$10   & 113$\pm$5   & 17$\pm$11        & 120$\pm$2 & 77.5 & 1\tabularnewline

Mrk\,1210 && 62.58$\pm$0.45  & 127$\pm$22  & 26$\pm$22   & 179$\pm$83    & 160.0 & 2 \tabularnewline

\textbf{Mrk\,1419} &&  49$\pm$0.7     & 159$\pm$2   &  20$\pm$1  & 117$\pm$84    & 40.2 & 3 \tabularnewline

\textbf{NGC\,2273} &&  153$\pm$4.6  & 95$\pm$1    & 32$\pm$5   & 83$\pm$3  & 63.3 & 3\tabularnewline 
UGC\,3193  &&   ---  & 175$\pm$0.2   & ---             & 167$\pm$4  & 177.4 & ---\tabularnewline
\hline
                                                                                                                                                                                                                                                 
\end{tabular}\par
\end{center}
\bigskip
\textbf{Notes}. Column 1: name of galaxy. The sources in bold hold clean maser disks (see Sect.\,\ref{sec:sample}).
Column 2: maser disk PAs and their uncertainties.
Column 3: the jet propagation direction, obtained from a linear fit on the position of the flux maxima in multi component sources (see Sect\,\ref{sec:orientation}). If there is only one component, 
the 5\,GHz continuum PA of this component represents the jet propagation direction.
Column 4: PA difference between the disk rotation axis (the normal vector to the disk) and jet propagation direction and the uncertainties.
Column 5: 33\,GHz larger scale PAs from \cite{kamali2017}.
Column 6: large scale galactic disk PA from HyperLeda.
Column 7: references for maser disk PA. (1)  Kuo et. al. (in prep.); (2) \citet{zhao2018}; (3) \citet{kuo2011}.
}                                                                       
\end{table*}

\section{Results}\label{sec:result}
Out of 18 sources observed, 5 were detected: Mrk\,0001, Mrk\,1210, Mrk\,1419, NGC\,2273 and UGC\,3193. Figure\,\ref{fig:images1} shows the 5\,GHz radio continuum maps. 
For all images, phase self-calibration was employed to improve the image quality, using a solution interval of 1 minute. 
For Mrk\,0001 the S/N was improved from 8 to 32, for Mrk\,1210 from 36 to 62, for Mrk\,1419 from 12 to 18, for NGC\,2273 from 56 to 61 and for UGC\,3193 from 8 to 36. 
For detected sources we fitted 2-dimensional Gaussians on the individual components using the task JMFIT in AIPS, in order to obtain source properties such as peak and integrated flux, position 
and its uncertainty (in case the continuum components are more complicated, multi-Gaussians are fitted over emission $> 3\,\sigma$). The result is presented in 
Table\,\ref{table:obs_properties}. 
For detected sources, in case a 
source has more than one component, the flux reported in Table\,\ref{table:obs_properties} is the 
total flux density which includes all the components\footnote{The radio luminosities are also obtained from the total flux densities unless mentioned otherwise.}. 
For positions reported in Tables\,\ref{table:calibrators} and \ref{table:obs_properties}, the uncertainties include the statistical uncertainty, the uncertainty in the phase calibrator position or the calibrator's size in case
the calibrator is resolved, and the systematic errors
which are estimated from $(\Psi^2+\rm beam^2)^{1/2}\times(1/[\rm S/R]+1/20)$, where $\Psi$ is the source major axis, beam is the beam size, and S/R is the signal-to-noise ratio \citep{white1997}.

\subsection{Continuum images and previous radio observations}\label{sect:individual sources}
 Here we review 
our maps and other radio observations available in the literature for the five detected sources.\\
\\
\textbf{- Mrk\,0001}\\
We observed compact radio emission in our 33\,GHz radio maps with a PA of $120\degree\pm2\degree$. In the C-band VLBA data we see extended emission, also in the northwest-southeast direction.
The maser disk has a PA of $6.4\degree\pm10\degree$ and its rotation axis is by $17\degree\pm11\degree$ misaligned from the jet propagation direction. The systemic masers are not observed in this source 
(Kuo et. al. in. prep.).
\\
\textbf{- Mrk\,1210}\\
As mentioned before, four of our VLBA detected sources were re-observed to obtain higher S/Ns. While for three sources morphologies are very similar in both observations,
for Mrk\,1210 the southeastern component was not detected in the data set from the first observations and the northwestern extended emission is less prominent in this data,
likely due to insufficient sensitivities.
A former 6\,cm VLBA observation of this galaxy has shown a similar structure as our re-observations which is shown in Fig.\,\ref{fig:images1}, except that the northwestern component is resolved 
into four sub-components \citep{middelberg2004}. The flux reported by \citet{middelberg2004} for the central component in Fig.\,\ref{fig:images1}
is the same as in our observations (5\,mJy). However the flux reported for the northwestern component is less than ours (9.1$\pm$1.1\,mJy compared to 13.7$\pm$0.4\,mJy). The southeastern component is considered
a tentative detection in the work of \citet{middelberg2004} and therefore no flux is reported for this component. Identifying the core of the galaxy from where the jet is emanating is
not easy since both northwestern and central components have steep spectra (\citet[][]{middelberg2004} reported a spectral index of $\alpha^{18\,\rm cm}_{6\,\rm cm}$=1.26 for the northwestern and $\alpha^{18\,\rm cm}_{6\,\rm cm}$=0.78 
  for the central component obtained from VLBI observations with matching beam sizes using the S $\propto$ $\rm \nu^{-\alpha}$ convention). The 
  maser configuration is consistent with a a rotating disk. However, only one single maser component is detected at the systemic velocity of the galaxy. 
  The high velocity masers also deviate slightly from an edge-on, flat, Keplerian disk model. This implies that if the
  masers arise from a disk, its inclination is 
  deviating $\sim$10\degree\, from an edge-on disk (inclination=100.68$^{+1.13}_{-0.90}$\degree), and there is potential warping or eccentricity in the disk \citep{zhao2018}.

Mrk\,1210 shows long-term variability in the X-rays and has changed from Compton-thick to Compton-thin over a period of approximately 17 years \citep{masini2017}. 
This variability was interpreted as a combination of the intrinsic variability of the AGN and presence of an obscuring cloud passing through the line of sight. 
\\
\textbf{- Mrk\,1419}\\
This galaxy poses a clean maser disk \citep{kuo2011}. \citet{sun2013} reported extended emission with a PA of $125\degree\pm10\degree$ from 20\,cm radio data. In our VLA maps we
see a weak extension in the southeast-northwest direction with a PA of 117\degree$\pm$84\degree. Gaussian deconvolution of our 5\,GHz map shows a compact component that
is not clearly resolved and may contain a weak extension with a PA of 159\degree$\pm$2\degree.
The outer part of the maser disk shows some degree of warping \citep{kuo2011}.
\\
\textbf{- NGC\,2273}\\
This galaxy poses a clean maser disk \citep{kuo2011}. NGC\,2273 was previously observed with the VLBA at C-band in 2004 by \citet{lal2004} and showed a compact radio core. \citet{anderson2005} observed this galaxy at 8.4\,GHz with the VLBA and 
detected radio emission extended along the northwest to southeast direction with a beam width of approximately 2\,mas $\times$ 1\,mas. The structure is similar to our map shown in the lower right panel of Fig.\,\ref{fig:images1}.
The distribution of the masers shows a hint of a warp in the disk \citep{kuo2011}.
\\
\textbf{- UGC\,3193}\\
In our 33\,GHz data, we see extended emission in the north-south direction, the same as what we observe in the C-band VLBA data.
The southern component is not resolved in the Gaussian deconvolution. The maser disk orientation and position is not yet known for UGC\,3193.

\subsection{A note on the non-detections}
\begin{figure}[h!]
 \includegraphics[width=250bp,height=210bp]{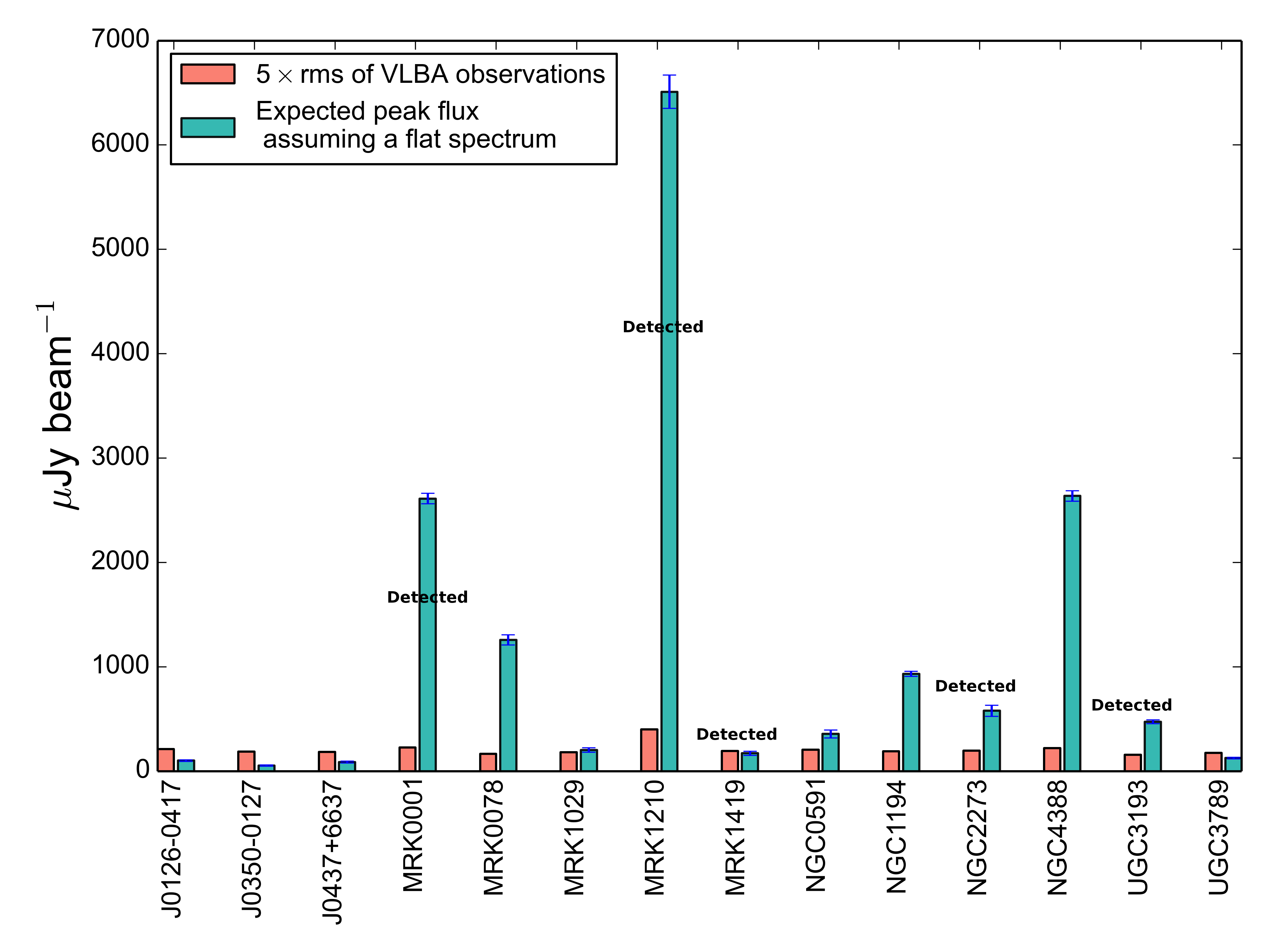}
 \caption{Expected flux densities in case of a flat radio spectrum compared to the rms noise of the maps. The 5 detected sources are labeled.}\label{fig:detections}
\end{figure}

About 72\% of the sources in our original sample of 18 sources were not detected. Considering that $\sim$70\% of the undetected 
sources were previously detected in our VLA observations and assuming a flat spectrum or rising flux densities
with decreasing frequency,
one might expect to detect these sources in the VLBA observations. 
Figure\,\ref{fig:detections} shows the expected flux at 5\,GHz (assuming a flat spectrum) and the 5$\times$rms noise of our VLBA maps (for a 5$\sigma$ detection limit of a single spatially unresolved
continuum component).
The five detected sources are among the brightest ones as shown in Fig.\,\ref{fig:detections}. However, there are similarly bright sources among the undetected ones.
Either these sources have inverted spectra, or extended emission is resolved out. Time variability of AGNs is another issue that could affect the
detection rates. 
We searched for trends of common properties among the detected sources, such as X-ray, IR, radio or maser luminosity limits, but did not succeed in finding any.
We report 3$\sigma$ upper limits for the undetected sources, rms of our maps, as well as hard X-ray and 33\,GHz fluxes in Table \ref{table:properties}.

\section{Discussion}\label{sec:discussion}
\subsection{Orientation of the jets with respect to the maser disks}\label{sec:orientation}
In NGC\,4258, the prototypical H$_2$O megamaser galaxy,
the rotation axis of the maser disk is aligned with the inner part of the extended jet \citep{cecil1992}.
Here we focus on identifying the  orientation of the jet with respect to that of the maser disk in our sample. 
Once again we emphasize that two of the detected sources, Mrk\,1419 and NGC\,2273, are among the clean maser disks (see Sect.\,\ref{sec:sample}), while the other three detected galaxies, Mrk\,0001, Mrk\,1210 and UGC\,3193,
are among those exhibiting a the non-clean disk, i.e., the masers might arise from either disks or outflows. In the following discussion,
we consider the possibility that the masers arise from the circumnuclear disk surrounding the SMBHs\footnote{For clarity, clean maser disks and non-clean maser disks are shown with different markers in our plots.}. 
Then for four out of the five detected galaxies 
in our sample, Mrk\,0001, Mrk\,1210, Mrk\,1419 and NGC\,2273, the 
maser disk orientation is known (see Sect.\,\ref{sec:result} for more information). As mentioned before, in our sample the maser disk is viewed edge-on ($\pm$10\degree) and the normal vector to the disk is in the plane of the sky. 
We do not apply any inclination corrections to the measured position angles (PAs). 
Instead we assume that the PA (measured east of north) of the radio continuum shows the PA of the elongation of the jet. In case that more than one component is detected, 
we assume that a line going through the central position of each component (obtained from Gaussian fits) is 
representing the jet propagation direction. The same convention for measuring the PAs (east of north) is used for all other reported PAs.  

\begin{figure*}[h!]
 \begin{center}
 \includegraphics[width=.405\textwidth]{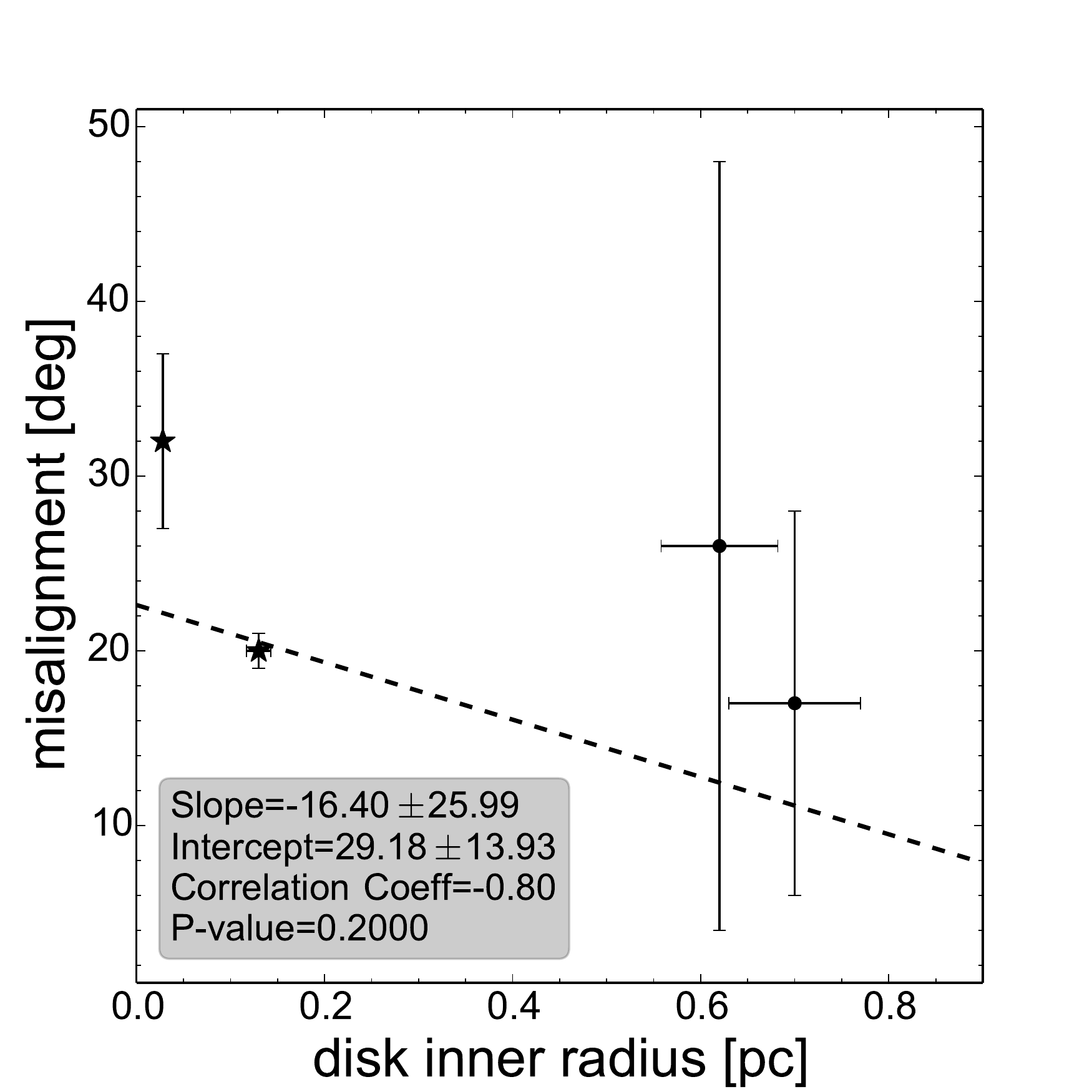}\hspace{0.5cm}\includegraphics[width=.38\textwidth]{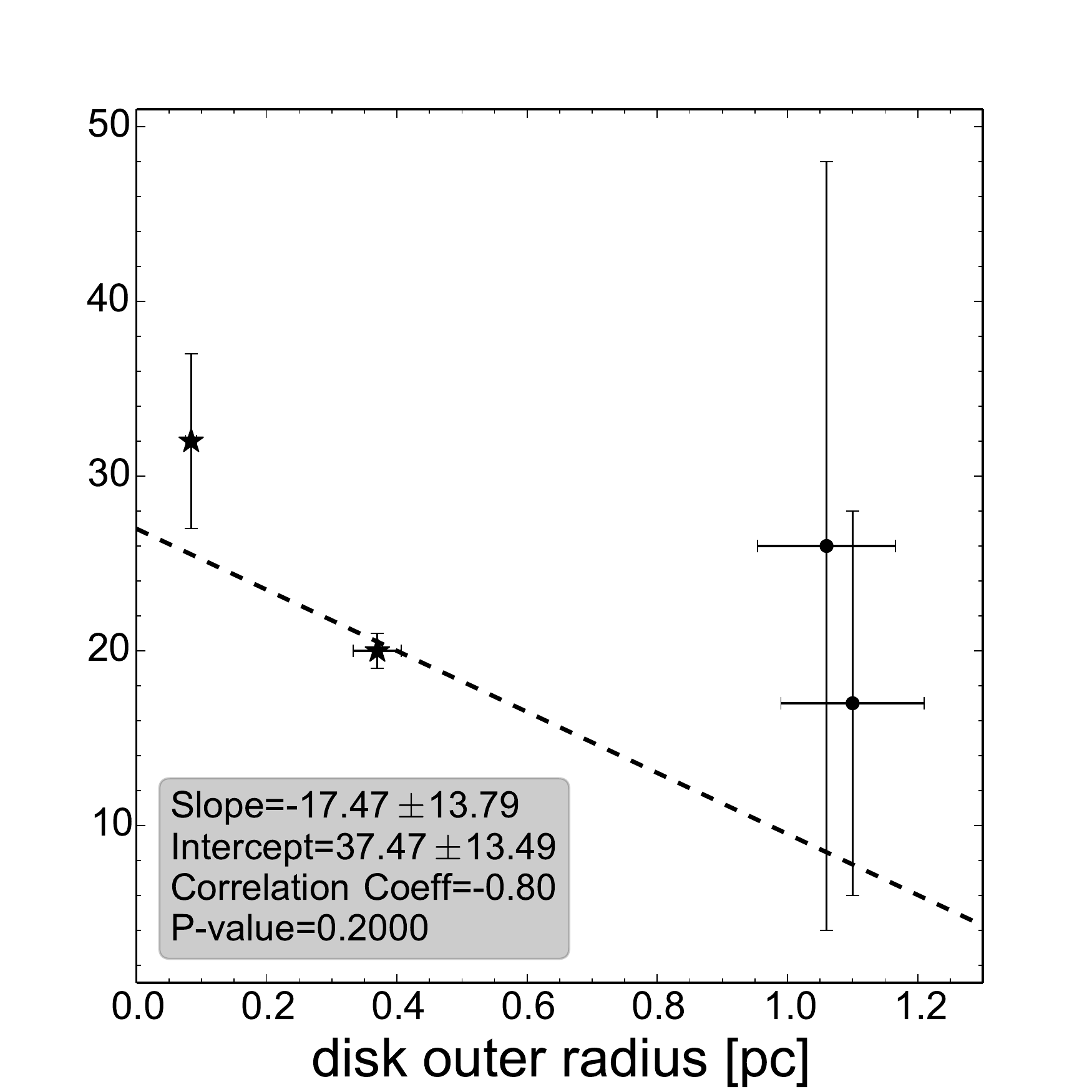}
 \end{center}
\caption{The misalignment of the jet and the normal to the maser disk, versus inner (left) and outer (right) 
radius of the maser disk. The clean maser disks are shown with star symbols and the non-clean disks with dots. }\label{fig:misalignmentradius}
\end{figure*}

\begin{figure*}[h!]
\begin{center}
  \includegraphics[width=480bp]{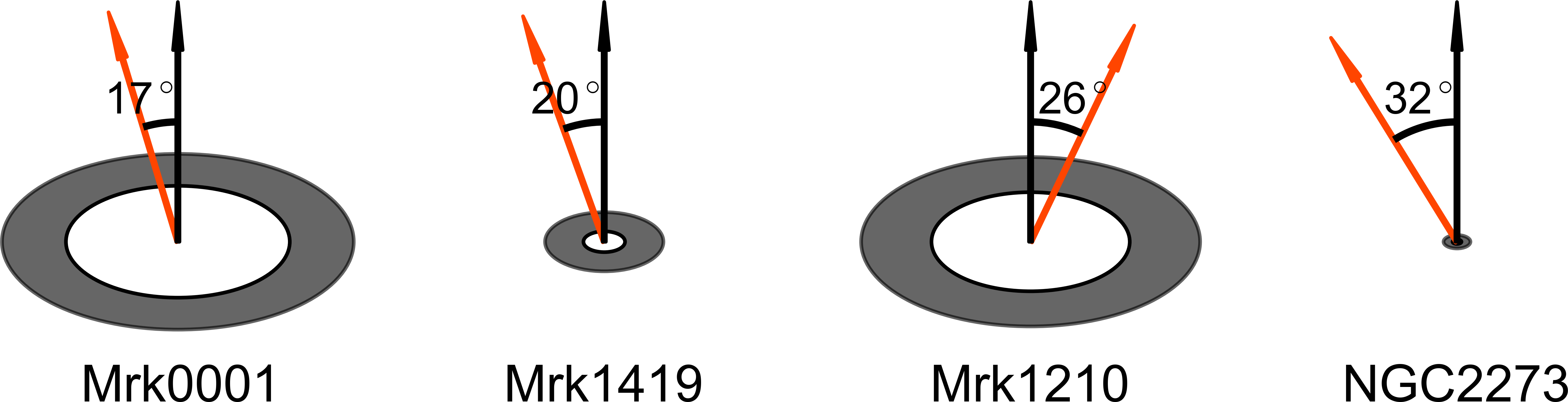}
  \caption{The jet (orange arrow) misalignment from the rotation axis (black arrow) of the maser disk (see also Table\,\ref{table:PAs}). The disk sizes are to scale with respect to each other.}\label{fig:jet-disknormal offset}
 \end{center}
 \end{figure*}

For all four sources, the jets are misaligned with respect to the normal vector to the maser disk.
However they are confined to a cone within $\sim$0.6 radians (32\degree) of the disks' normal. 
Table\,\ref{table:PAs} presents
PAs obtained in this work and from the literature and Fig.\,\ref{fig:PAs} shows a comparison between 
the PAs of different scale disks and the radio continuum.
For the null hypothesis that the jets have random distribution in space (which means they do not care about the maser disk), 
the probability that we observe these four measured PA offsets
is 0.009, indicating that the jet orientations are not random with a 99.1\% confidence level (see Appendix\,\ref{appendix:stat} for more details).
On the other hand, the PA offset may be related to the outer and inner radii of the disk: the larger the disk size, the smaller the offset (see Figs.\,\ref{fig:misalignmentradius} and \ref{fig:jet-disknormal offset}).
We note, however, that our conclusion is based on a sample of four sources only (where two of them are non-clean disks), so that a study of a larger sample is needed to verify or falsify our finding.

\begin{figure*}[h!]
\begin{center}
  \includegraphics[width=.7\textwidth]{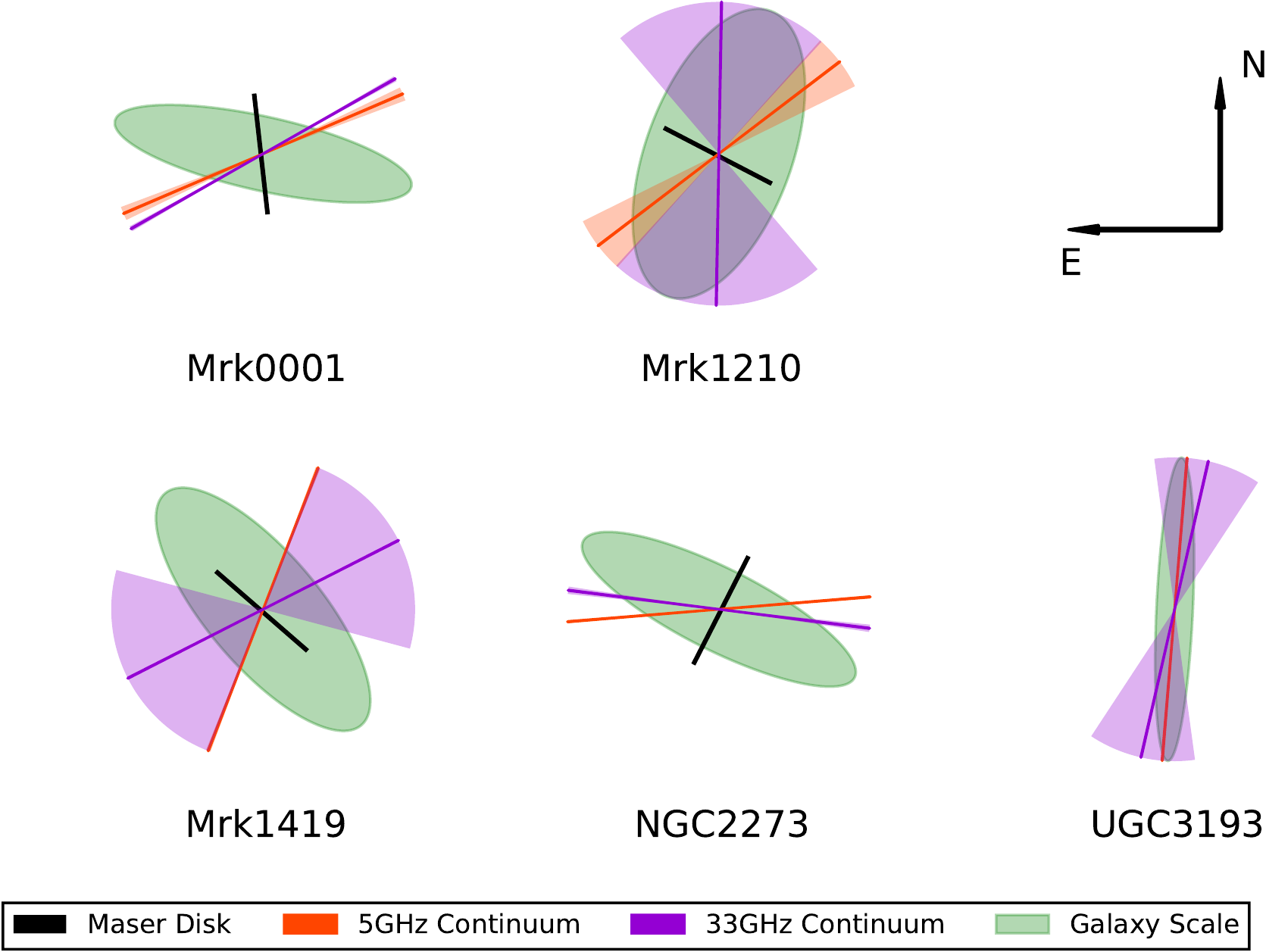}\label{fig:PAs}
\end{center}
\caption{Position Angles (PAs) on different scales. The galaxy scale PA is shown as a green ellipse. The minor axis of the ellipse represents the inclination of the galaxy, i.e., the broader the minor axis, the lower the inclination.
The 5\,GHz VLBA and 33\,GHz VLA radio
continuum PAs are shown with orange and violet solid lines respectively, where the uncertainties in the PA of the radio continuum are shown as shaded wedges. 
The maser disk is shown with a black solid line. For UGC\,3193 detailed maser images are not yet
available. The sizes of the ellipses, lines and wedges do not represent correct relative scales.}\label{fig:PAs}
\end{figure*}
\subsection{Multi scale position angle}
The PAs of disks on different scales are available for three of the VLBA-detected sources \citep{greene2013, pjanka2017}. Therefore 
we can compare the alignment of the jet with the angular momentum orientation of the disks on different scales.
Our VLBA radio continuum in Mrk\,1210 is misaligned by ${\sim}26\degree$ with the maser disk's normal. However, it is misaligned with the nuclear region of the galaxy (with size of 170\,pc) by only ${\sim}8\degree$.
The radio continuum orientation is displaced by ${\sim}57\degree$ from the angular momentum orientation of the large scale disk. 
For NGC\,2273, the VLBA radio continuum is not aligned with any of the rotation axes: ${\sim} 32\degree$ from the maser disk's normal, ${\sim} 36\degree$ from the angular momentum orientation of the nuclear 150\,pc scale disk
and ${\sim}58\degree$ from the normal vector to the galaxy's large scale disk.
In Mrk\,1419, which has likely experienced a merger or some interaction \citep{lasker2016},
the maser disk and the nuclear region with a size of 270 pc are aligned and the jet is ${\sim}20\degree$ from the angular momentum direction of this disk, while the 
galaxy's large scale disk has a normal vector that is ${\sim} 29\degree$ away from the jet direction.
\subsection{Radio continuum positions relative to the maser positions}
Compact radio emission is detected in the nuclei of most galaxies \citep[][and references therein]{herrnstein1997}. Since the centers of the maser disks are the gravitational centers of the nuclear regions,
we expect the centers of the maser disks to coincide with their radio continuum core representing the nucleus of the galaxy.
For three sources among the detected ones, Mrk\,0001, Mrk\,1419 and NGC2273, the maser positions are available. For our radio continuum data the central position from the Gaussian fit of the component with the highest brightness temperature (if there are more than one)
is considered as the position of the nuclear source 
(see Sect.\ref{sec:index}). In Table\,\ref{table:separation} we present the spatial offsets of the maser disk with respect to the radio continuum. 
For Mrk\,1419 and NGC\,2273, the maser
positions are from VLA observations. Within the uncertainties, the maser positions for all three sources are consistent with those of the radio continuum. For Mrk\,0001, however, 
in our second observations the phase calibrator was resolved with a major axis of $\sim$\,15 mas and therefore we report the position offset with respect to the continuum position in our first observations which 
has a higher position accuracy.
The proper motion of the galaxies can be neglected. In case of a speed of $1000\,\mathrm{km}\,\mathrm{s}^{-1}$ in the plane of the sky and a distance $D=50\,\mathrm{Mpc}$, the expected shift is only 4\,$\mu$arcsec\,yr$^{-1}$.

\subsection{Spectral indices and brightness temperatures}\label{sec:index}
We obtained the spectral indices for the detected sources between 33\,GHz and 5\,GHz. They are presented in Table \ref{table:obs_properties}. 
The indices were obtained assuming a power-law dependence for the fluxes (S $\propto$ $\rm \nu^{-\alpha}$). We should note that the VLA observations have 
linear beam sizes about 100 times those of the VLBA observations. The spectral indices have a mean of $\alpha$=0.21$\pm$0.20.

We also obtained the brightness temperatures ($\rm T_B$), using  
\begin{equation}
 T_b=\frac{2\,ln(2)}{k_B\, \pi}\frac{\,\lambda^2\,S_{tot}}{\,\theta_{maj}\,\theta_{min}}
\end{equation}
where $k_B$ is the Boltzmann constant, $\rm S_{tot}$ is the total flux density, $\lambda$ is the wavelength of observation, and $\theta_{maj}$ and $\theta_{min}$ are the deconvolved sizes of the major and minor axes of the source. When the component is unresolved, the beam size is used as an upper-limit for the source size. Therefore the corresponding $\rm T_B$ is a lower limit. The obtained brightness temperatures are reported in Table\,\ref{table:obs_properties}. For sources with more than one component, we report the measurement for the component with highest $\rm T_b$, which we identify as the putative core of the AGN. It should be noted that we do not know the spectral indices for the individual components, therefore the brightness temperatures are the only tool by which we can tentatively define the AGN's core.
All sources have $\rm T_B > 10^6$, consistent with non-thermal emission.

\begin{table}[t]
\caption {Separation of the maser disk from the continuum sources.} \label{table:separation}
\begin{center}
\begin{tabular}{l c c cl } 
\hline \hline
&  &\tabularnewline
\textbf{Galaxy}  &  \textbf{$\delta$ RA } &\textbf{ $\delta$ Dec} & \textbf{Reference } \tabularnewline
 & (mas) & (mas)

\\
\hline
\\
Mrk0001		& 4.5$\pm$1.0    &  0.4$\pm$1.0	    & 1 \tabularnewline 

Mrk1419		& 1.5$\pm$10.0    & 1.5$\pm$10.0	& 2\tabularnewline
NGC2273		& 22.5$\pm$10.0	  & 3.3$\pm$10.0    & 2  \tabularnewline
       
\\
\hline
\end{tabular}\par
\end{center}
\bigskip
\textbf{Notes}. Column 1: name of galaxy.
Column 2: Separation of the maser disk from the center of the 5\,GHz nuclear continuum source and 
the uncertainties. For Mrk\,0001, the separation is for our first observation and with respect to the continuum source in the 
northwest of the image.
Column 3: References for maser disk positions. (1) Kuo  et al. in prep.; (2) \citet{kuo2011}.
\end{table}
\begin{figure*}[h!]
 \includegraphics[width=.5\textwidth]{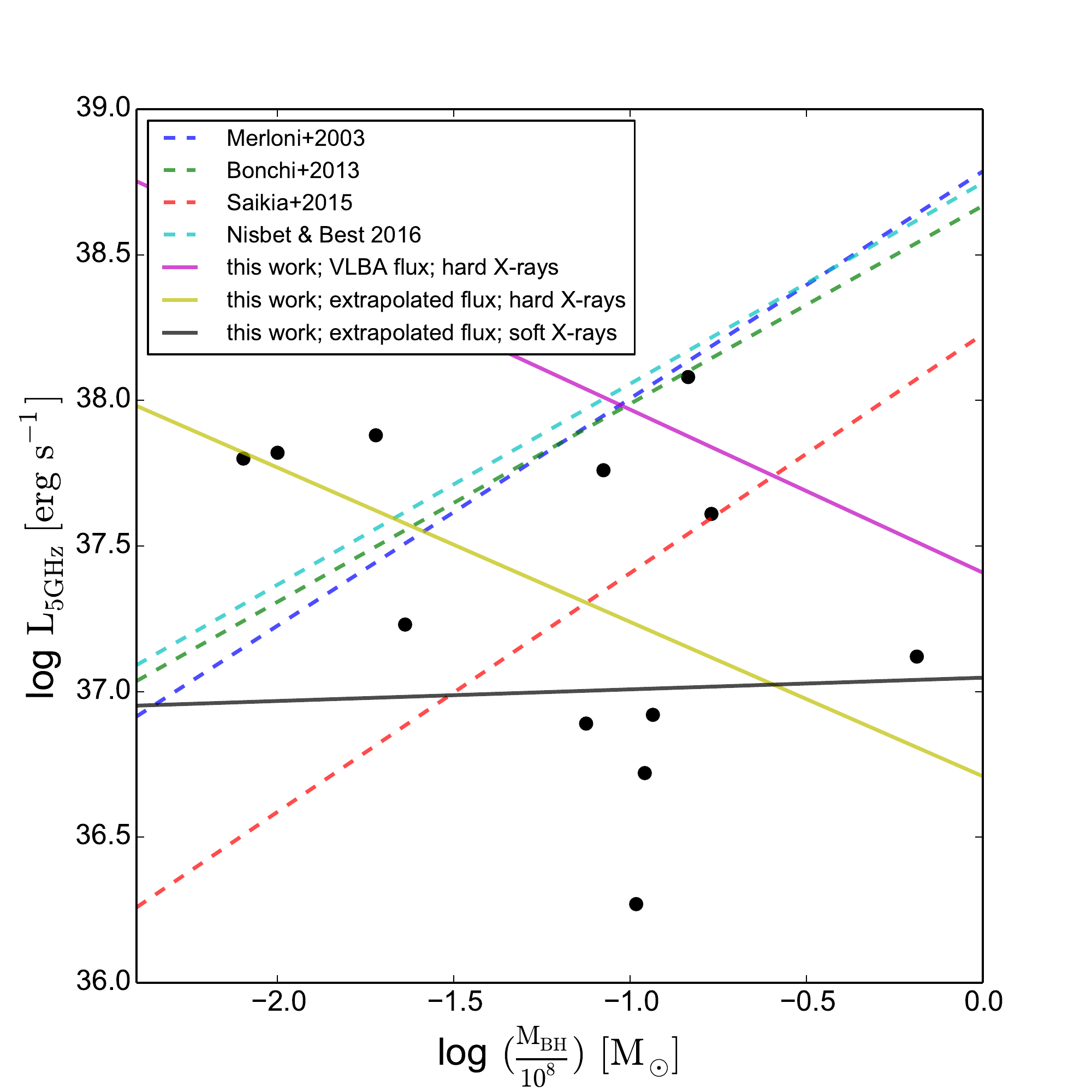} \includegraphics[width=.5\textwidth]{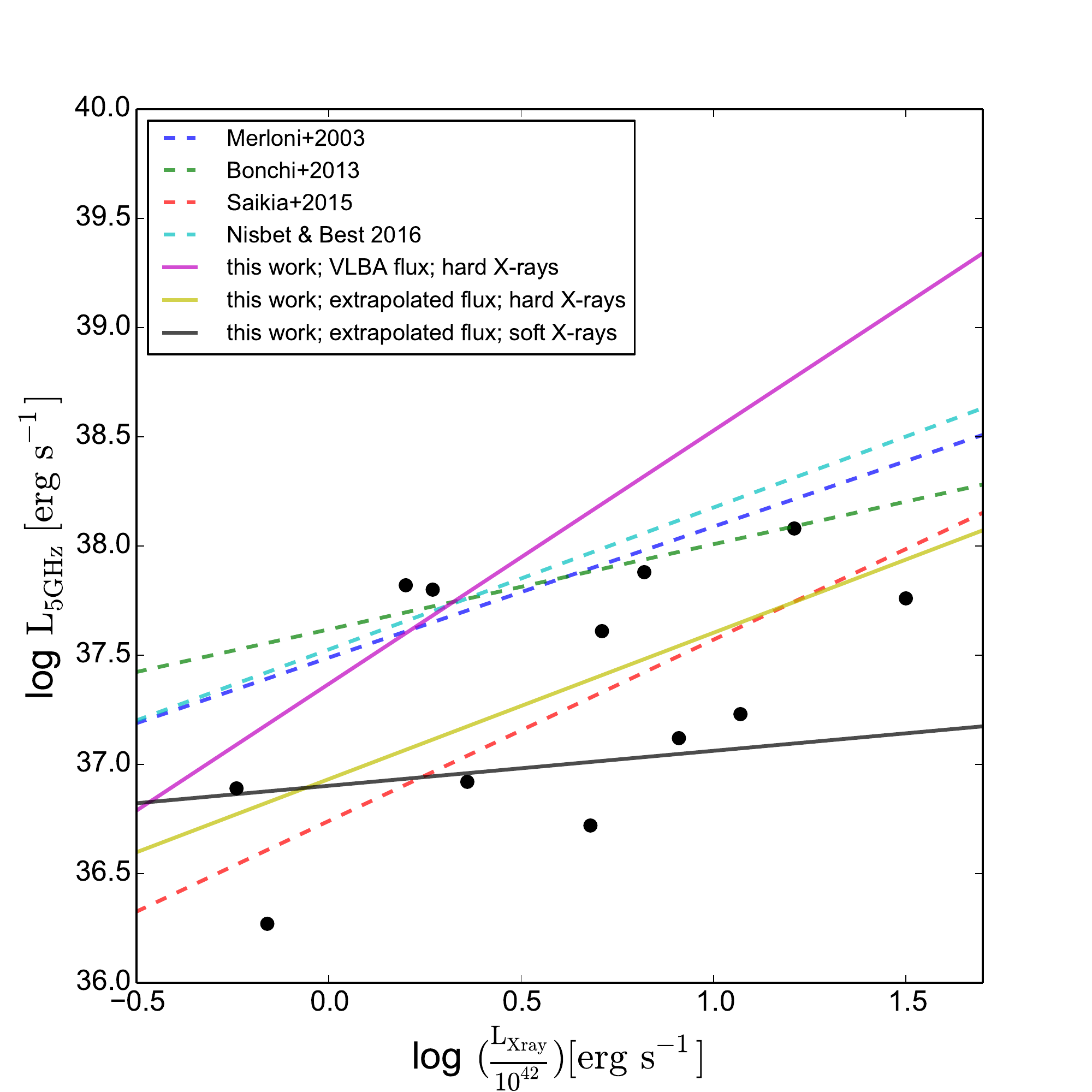}
 \caption{Parameters of the fundamental plane of the black hole activity. Left panel: 5\,GHz radio luminosity versus black hole mass. 
 The data points are the extrapolated 5\,GHz fluxes obtained from the 33\,GHz data \citep{kamali2017}, using a spectral index of 0.21 (see Sect.\,\ref{sec:fundamental-plane}).
 Dashed lines present $\xi_{\rm M}$ available in the literature (see equation\,\ref{eq:fundamental-plane}), assuming a constant X-ray luminosity 
 (mean value of our data)
 and the solid lines present $\xi_{\rm M}$ obtained in this work.
 Right panel: 5\,GHz radio luminosity versus X-ray luminosity. Dashed lines present $\xi_{\rm X}$ available in the literature, assuming a constant black hole mass (mean value of our data)  and the solid
 lines present $\xi_{\rm X}$ obtained in this work. }
\label{fig:fp}
\end{figure*}
\subsection{Radio luminosity versus other properties of the galaxies }
\subsubsection{Radio luminosity vs. X-ray luminosity and the SMBH mass: fundamental plane of black hole activity}\label{sec:fundamental-plane}
The fundamental plane (FP) of black hole activity is an empirical relation between the radio continuum luminosity, X-ray luminosity and black hole (BH) mass. 
This relation holds for 10 orders of magnitude in radio luminosity and ranges from stellar mass BHs (in X-ray binaries) to SMBHs, supporting the idea that jet physics 
is scale invariant \citep{merloni2003}. 
For many years there have been efforts to understand the accretion mechanism in the SMBHs and how the accretion rate is related to
the SMBH mass and the radio and X-ray luminosities. Studies have suggested that the radio luminosity depends on both SMBH mass and accretion rate of material
onto the black hole \citep{falcke1995a}. The SMBH mass can be determined observationally, and the accretion rate can be inferred through 
measuring the luminosity at those frequencies
at which the accretion process dominates. For example, the accretion flow can be inferred from the X-ray luminosity since the X-ray emission is produced 
by the accretion of material onto the SMBHs. The scaling of the X-ray luminosity
with the accretion rate depends on the physics of the disk, e.g., whether it is radiatively efficient or not 
\citep[a quantitative description of these models is beyond the scope of this paper; see][and references therein]{merloni2003}. 
The X-ray emission could also be produced by the inverse 
Compton effect involving synchrotron radiation.
Summarizing everything, we expect the core radio luminosity to be correlated with the X-ray luminosity and also with the black hole 
mass. 

The following plane equation describes the FP of BH activity:
\begin{equation}\label{eq:fundamental-plane}
$log$\,L_R\,[$erg$\,$s$^{-1}]= \xi_{\rm X}\,$log$\,L_X\,[$erg$\,$s$^{-1}]+ \xi_{\rm M}\,$log$\,M\,[$M$_{\odot}] + b_R 
\end{equation}
where $L_{R}$, $L_{X}$ and $M$ are the radio luminosity, X-ray luminosity
and the BH mass respectively, $b_R$
is the zero intercept, and $\xi_{\rm X}$ and $\xi_{\rm M}$ are the linear regression coefficients.
The coefficients in Eq.\,\ref{eq:fundamental-plane} depend on the accretion efficiency, electron spectral index, viscosity of the disk and initial conditions at the base
of the jet such as dependence of magnetic energy density on the SMBH mass and accretion rate.
Others have followed the lead of \citet{merloni2003} and investigated the FP using different subsamples \citep[e.g.,][]{saikia2015,koerding2006,gueltekin2009,bonchi2013,nisbet2016}.
All these studies, fit  similar coefficients, which is somewhat surprising since the SMBH mass determinations differ from sample to sample. We note that a sample for which accurate
BH mass measurements have been obtained from edge-on megamaser disks has not yet been analyzed.

Adopting X-ray luminosities from Litzinger et al. (in. prep), obtained using data from the
\textit{Swift}/BAT instrument in a range of 20-100 keV (see Table\,\ref{table:properties}), the black hole masses from the MCP and radio luminosities from this work,
we fitted the same function to our data using the least square method. The fitting parameters are shown in the first row of Table\,\ref{table:fundamental-plane}.
We considered increasing the number of sources for this analysis by extrapolating the 5\,GHz luminosities from the
33\,GHz radio luminosities, using the mean of the indices between our 33\,GHz and 5\,GHz observations (see Table.\,\ref{table:obs_properties}).
This sample with extrapolated fluxes includes 12 sources that have known SMBH masses. The result of the fit 
is presented in the second row of Table\,\ref{table:fundamental-plane}.
We should note that the mentioned studies that probe the FP use soft X-ray (2-10\,keV) data and this can cause a discrepancy
to our result. We also tried to fit Eq.\ref{eq:fundamental-plane} to a sample of 5 sources for which we have soft X-ray data \citep{masini2016} 
and extrapolated 5\,GHz fluxes. The fitting parameters are shown in the third row of Table\,\ref{table:fundamental-plane}.
We present the fitting parameters for the FP in Fig.\,\ref{fig:fp}. A selection of parameters reported in the literature is also presented with dashed lines.  
As seen in Fig.\,\ref{fig:fp}, in our sample the black hole mass decreases with increasing radio luminosity. 
This is opposite to what is found in other studies
and could be responsible for the significant difference between
this work and other studies for the obtained value of $\xi_{\rm M}$.
On the other hand, galaxies with H$_2$O megamaser disks may not follow the well-known scaling relation between the BH mass 
and the stellar dispersion velocity of the galaxies. They may show an offset with respect to the early-type
galaxies, while other spiral galaxies with non-maser dynamically determined BH masses do not show such an offset \citep{greene2016}. This could be because
the maser disks are preferentially formed in spirals with particularly low mass nuclear sources, or because the non-maser dynamical measurements
are missing the low-mass end of the BH distribution due to their too small spheres of influence. It should be noted that these two above mentioned interpretations are only two possible scenarios among other possible ones.

\subsubsection{Radio luminosity vs. [OIII] luminosity}\label{sect:OIII} 
\begin{figure}[h!]
\begin{center}
  \includegraphics[width=0.36\textwidth]{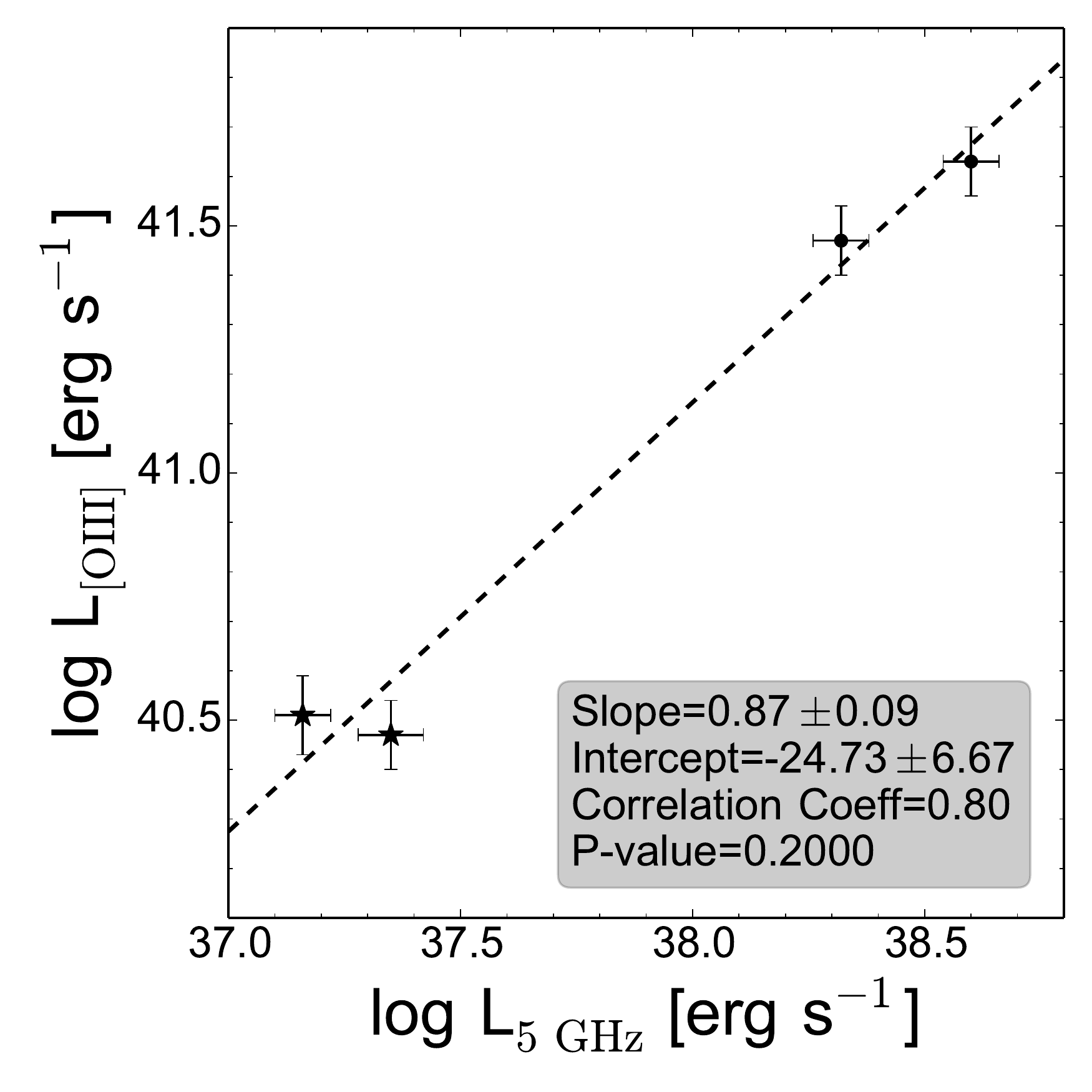}
  \caption{5007\,$\AA$ [OIII] versus 5\,GHz luminosity. The slope and the intercept of the fitted line,
correlation coefficient and the P-value of the rank correlation test are shown in the plot. The clean maser disks are shown with star symbols and the non-clean disks with dots. }\label{fig:OIII-radio}
 \end{center}
 \end{figure}

Previous studies have shown a strong correlation between the optical narrow emission-line luminosity and the radio luminosity of radio galaxies \citep[e.g.,][]{baum89, rawlings91}. 
This correlation
is valid over four to five orders of magnitude for both radio and line luminosities and 
could be the result of a common energy source for optical line and radio emissions, suggesting that the luminosity depends on the properties of the central engine. It is also possible
that both optical line and radio luminosities depend on a third parameter, e.g., the amount of cold gas present on kpc scales \citep{baum89}.
A correlation was reported by \citet{baum89} between the H$\alpha$+[N II] line luminosity, ${\rm L}_{\rm line}$,
and total radio luminosity at the VLA scales for a sample of radio galaxies where L$_{\rm line}$ is proportional to L$_{\rm radio}^{0.73}$.
Other studies have compared the radio core luminosity and the luminosities from the narrow line region, where the core is defined as the unresolved component with highest flux density in the radio observations.
\citet{buttiglione2010} reported
 L$_{\rm [OIII]} \propto$ L$_{\rm core}^{0.75}$ for a sample of high excitation galaxies and \citet{baldi2018} reported L$_{\rm [OIII]} \propto$ L$_{\rm core}^{0.35\pm0.20}$ for a sample of LINER galaxies.
Therefore, we check for a possible linear correlation between the optical emission-line luminosities and radio luminosities in our sample.
We obtained the 5007\,$\AA$[OIII] luminosities from NED taken by the SDSS. 
Four of our five detected sources, Mrk\,0001, Mrk\,1210, Mrk\,1419 and NGC\,2273, have [OIII] luminosities. Using a Spearman rank correlation test, we find a correlation between 5\,GHz (derived from VLBI) and
the [OIII] luminosities, L$_{\rm [OIII]} \propto$ L$_{\rm 5\,GHz}^{0.87\pm0.09}$. The correlation coefficient is 0.80, and the P-value, indicating that the likelihood that parameters are unrelated, is 0.2
(see Fig.\,\ref{fig:OIII-radio}). While this suggests reasonable agreement with previously mentioned studies, we note the result is based on very few sources only.
We also note that the correlation of radio luminosity  with emission-line luminosity extends over five orders of magnitude in L$_{\rm radio}$ and four orders of magnitude in L$_{\rm line}$,
i.e., there is $\sim$ one order of magnitude scatter in the radio luminosity for a given line luminosity, suggesting that other criteria such as environment
or different accretion modes play a role in determining the radio and optical emission-line luminosities of these galaxies \citep{baum89}.

\begin{table*}[t]
\caption {Fit parameters of fundamental plane for black hole activity.} \label{table:fundamental-plane}
\begin{tabular}{l l l l} 
\hline \hline
& \tabularnewline
\textbf{$\xi_{\rm X}$}  &  \textbf{$\xi_{\rm M}$} &  \textbf{$b_R$} & \textbf{Notes}\tabularnewline
\\
\hline
\\
 1.16$\pm$1.11 & -0.56$\pm$0.97 & -7.61$\pm$45.13 & this work, hard X-rays, 5\,GHz VLBA fluxes, 5 sources\tabularnewline
  0.67$\pm$0.20 & -0.53$\pm$0.19 & 12.27$\pm$8.18 & this work, hard X-rays, extrapolated 5\,GHz fluxes using $\alpha$=0.21, 12 sources\tabularnewline
  0.16$\pm$3.3  & 0.04$\pm$5.6 & 30.1$\pm$141 & this work, soft X-rays, extrapolated 5\,GHz fluxes using $\alpha$=0.21, 5 sources \tabularnewline
 \\
\hline
\end{tabular}\par
\bigskip
\textbf{Notes}. Column 1: coefficient of X-ray luminosity in Eq.\ref{eq:fundamental-plane}.
Column 2: coefficient of SMBH masses in Eq.\ref{eq:fundamental-plane}.
Column 3: intercept of Eq.\ref{eq:fundamental-plane}.
Column 4: notes. Extrapolated 5\,GHz fluxes are obtained using spectral index 0.21 between 33\,GHz VLA observation and 5\,GHz VLBA observations assuming a power law dependence of S $\propto$ $\rm \nu^{-\alpha}$.
\end{table*}

 \subsubsection{Radio luminosity vs. maser disk radius}
\begin{figure}[h!]
\begin{center}
\includegraphics[width=0.36\textwidth]{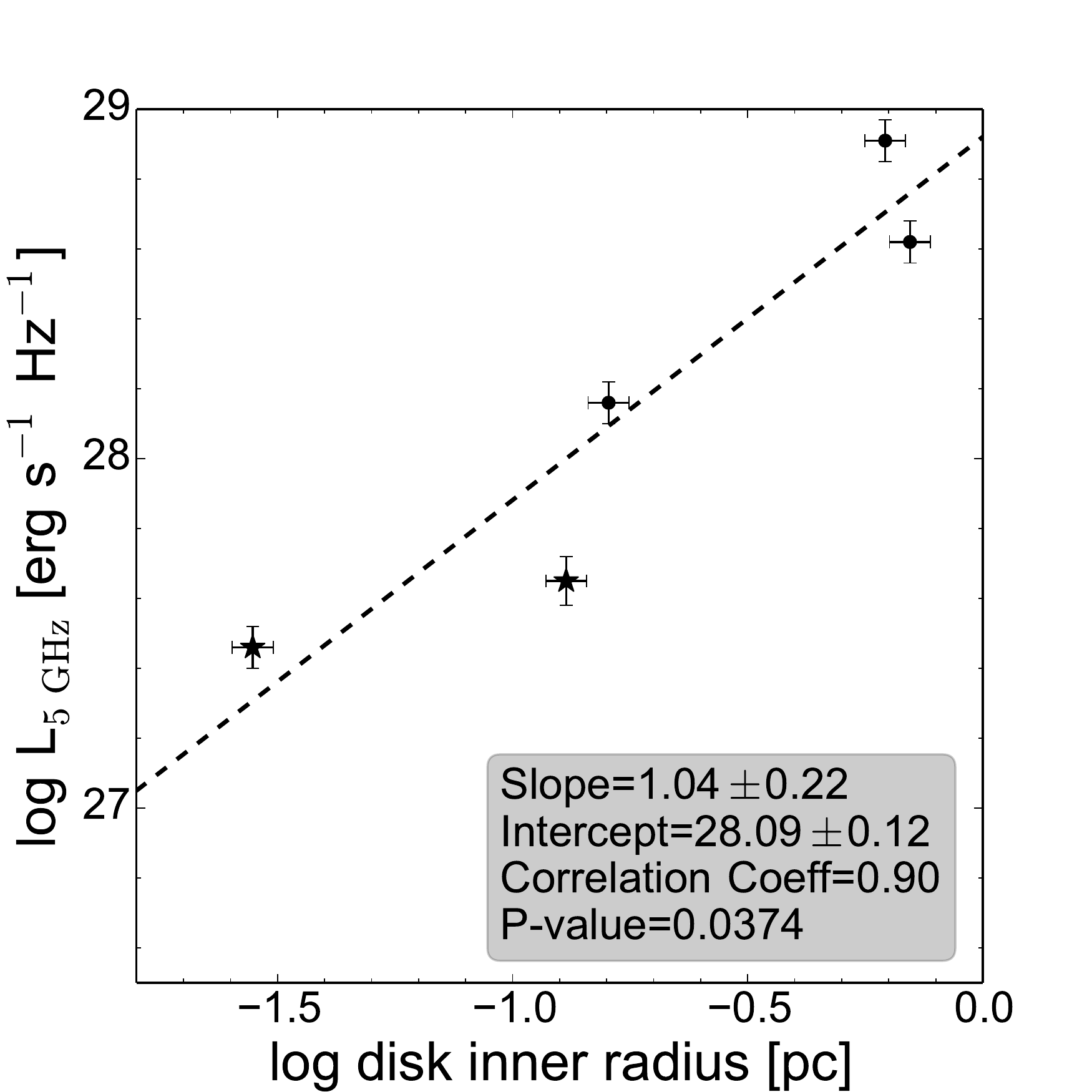}
\includegraphics[width=0.35\textwidth]{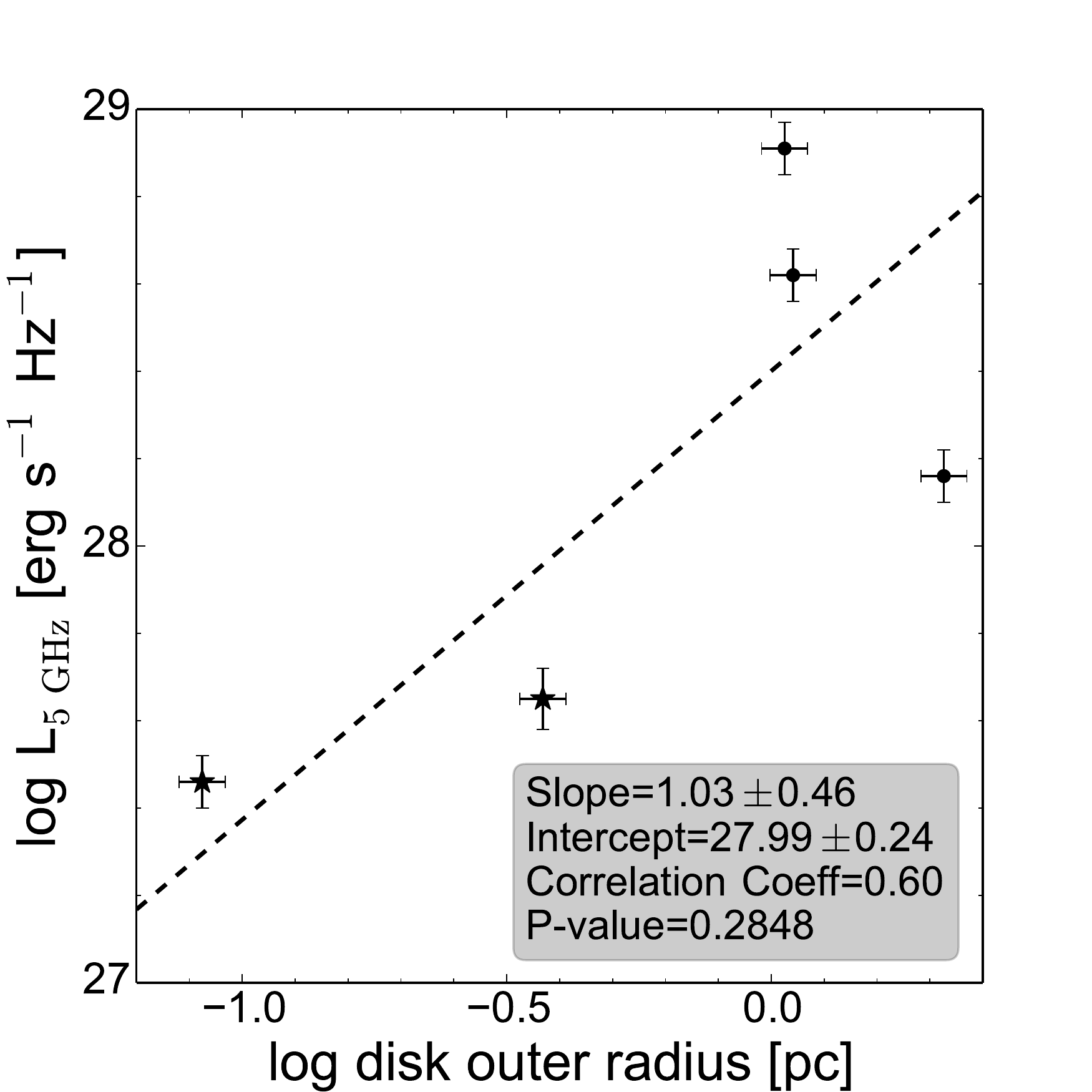}
\caption{5\,GHz luminosities versus the inner radius (upper panel) and the outer radius (lower panel) of the maser disk. The slope and the intercept of the fitted line,
correlation coefficient and the P-value of the rank correlation test are shown in the plot. The clean maser disks are shown with star symbols and the non-clean disks with dots.}\label{fig:lum_radius}
\end{center}
\end{figure}

 In our pilot project \citep{kamali2017} we demonstrated a correlation between the disk's inner and outer radii and the 33\,GHz radio luminosity of galaxies with H$_2$O masers in their nuclear disks
 (i.e., both inner and outer radii of the disk increase as the radio luminosity increases).
 In this work, 
 we find a similar correlation for inner and outer disk radius versus the 5\,GHz luminosities (Fig.\,\ref{fig:lum_radius}). 
 Correlations between the maser disk size and luminosity at other frequencies have been investigated formerly. For example, \citet{gao2017} reported a correlation between
 the infrared (IR) luminosity and the maser disk's outer radius. This correlation exists because the gas temperature should be more than the minimum
 kinetic temperature of $\sim$\,400\,K for masing to occur. The H$_2$O molecules are mixed with dust and the dust 
 temperature depends on the bolometric luminosity. Therefore the outer radius of the maser disk also depends on the bolometric luminosity of the AGN.
 In Sect.\,\ref{sect:OIII} we showed a correlation between the 5\,GHz luminosity and the [OIII] luminosity \citep[which is a proxy for the AGN's bolometric luminosity, e.g.,][]{bassani1999,heckman2005}.
Thus a correlation between the outer radius of the maser disk and 5\,GHz luminosities is expected.
Furthermore, the pressure outside the outer disk is less than the critical pressure which permits the existence of the molecules \citep{neufeld1994} and this critical pressure depends on the
X-ray luminosity \citep[for detailed analysis see][]{neufeld1995}.
Although a correlation between the X-ray luminosity and the maser disk size is expected, former studies did not find such a correlation \citep{kamali2017}.
High energy electrons are responsible for both synchrotron radio emission and the synchrotron self Compton X-ray radiation \citep{falcke1995a}. 
However, our sources are Compton thick and a misestimate of the column densities can affect the estimation of the intrinsic unobscured X-ray luminosities. Radio emission
does not suffer from extinction and might be a better tool to estimate the impact of these high energy electrons that are confining the disk.


\section{Summary}\label{sec:sum}
We observed a sample of 18 LLAGNs with the VLBA at C-band (6\,cm) and detected 5 sources: Mrk\,0001, Mrk\,1210, Mrk\,1419, NGC\,2273 and UGC\,3193.
We aimed to measure the orientations of the putative jets with respect to those of the maser disks. Excecpt for UGC\,3193, for the other four detected sources, the maser disk orientation is known.
All four sources have jet directions that are misaligned with the disk's normal by < 0.6 radian. 
For the null hypothesis that the jets have random distribution in space, 
the probability that all four sources have a misalignment of  < 0.6 radian
is 0.009, indicating that the jet orientations are not random.
The PA differences seem to be related to the disk size: the larger the disk size, the more aligned is the radio continuum to the disk's normal.
A tight correlation with correlation coefficient of 0.87 is observed between the 5007\,$\AA$[OIII] line luminosity and the 5\,GHz radio continuum luminosity.
The disk's inner and outer radii show a linear correlation with the 5\,GHz continuum luminosity: the larger the inner radius of the maser disk the brighter is the central region of the galaxy at 5\,GHz.
Note, however, that all these results are based on a very small sample of sources.
Further observations and maps of maser disks and of the nuclear regions of maser host galaxies will provide a larger sample for a more detailed investigation on the alignment of the accretion disk and jets in the galaxies where both phenomena are
directly observable.


\begin{acknowledgements}
F.K. would like to thank the anonymous referee for the critical and constructive feedback. F.K. is also grateful to Gisela Noemi Ortiz-Leon for her constructive comments and discussions on the data reduction process. This work made use of the NASA/IPAC extragalactic Database (NED), which 
is operated by thr Jet Propulsion Laboratory, California Institute of Technology, under contract with NASA. We further acknowledge the usage of the Kapteyn Package \citep{KapteynPackage}, the HyperLeda database 
(http://leda.univ-lyon1.fr) and the SAO/NASA ADS Astronomy Abstract Service (http://adsabs.harvard.edu) and the National Radio Astronomy Observatory 
which is a facility of the National Science Foundation operated under cooperative agreement by Associated Universities, Inc.
\end{acknowledgements}

\bibliography{biblio}

\begin{thebibliography}{48}
\expandafter\ifx\csname natexlab\endcsname\relax\def\natexlab#1{#1}\fi

\bibitem[{{Anderson} \& {Ulvestad}(2005)}]{anderson2005}
{Anderson}, J.~M. \& {Ulvestad}, J.~S. 2005, \apj, 627, 674

\bibitem[{{Baldi} {et~al.}(2018){Baldi}, {Williams}, {McHardy}, {Beswick},
  {Argo}, {Dullo}, {Knapen}, {Brinks}, {Muxlow}, {Aalto}, {Alberdi}, {Bendo},
  {Corbel}, {Evans}, {Fenech}, {Green}, {Kl{\"o}ckner}, {K{\"o}rding}, {Kharb},
  {Maccarone}, {Mart{\'{\i}}-Vidal}, {Mundell}, {Panessa}, {Peck},
  {P{\'e}rez-Torres}, {Saikia}, {Saikia}, {Shankar}, {Spencer}, {Stevens},
  {Uttley}, \& {Westcott}}]{baldi2018}
{Baldi}, R.~D., {Williams}, D.~R.~A., {McHardy}, I.~M., {et~al.} 2018, \mnras,
  476, 3478

\bibitem[{{Bassani} {et~al.}(1999){Bassani}, {Dadina}, {Maiolino}, {Salvati},
  {Risaliti}, {Della Ceca}, {Matt}, \& {Zamorani}}]{bassani1999}
{Bassani}, L., {Dadina}, M., {Maiolino}, R., {et~al.} 1999, \apjs, 121, 473

\bibitem[{{Baum} \& {Heckman}(1989)}]{baum89}
{Baum}, S.~A. \& {Heckman}, T. 1989, \apj, 336, 702

\bibitem[{{Bonchi} {et~al.}(2013){Bonchi}, {La Franca}, {Melini}, {Bongiorno},
  \& {Fiore}}]{bonchi2013}
{Bonchi}, A., {La Franca}, F., {Melini}, G., {Bongiorno}, A., \& {Fiore}, F.
  2013, \mnras, 429, 1970

\bibitem[{{Braatz} {et~al.}(2010){Braatz}, {Reid}, {Humphreys}, {Henkel},
  {Condon}, \& {Lo}}]{braatz2010}
{Braatz}, J.~A., {Reid}, M.~J., {Humphreys}, E.~M.~L., {et~al.} 2010, \apj,
  718, 657

\bibitem[{{Buttiglione} {et~al.}(2010){Buttiglione}, {Capetti}, {Celotti},
  {Axon}, {Chiaberge}, {Macchetto}, \& {Sparks}}]{buttiglione2010}
{Buttiglione}, S., {Capetti}, A., {Celotti}, A., {et~al.} 2010, \aap, 509, A6

\bibitem[{{Cecil} {et~al.}(1992){Cecil}, {Wilson}, \& {Tully}}]{cecil1992}
{Cecil}, G., {Wilson}, A.~S., \& {Tully}, R.~B. 1992, \apj, 390, 365

\bibitem[{{Falcke} \& {Biermann}(1995)}]{falcke1995a}
{Falcke}, H. \& {Biermann}, P.~L. 1995, \aap, 293, 665

\bibitem[{{Gao} {et~al.}(2017){Gao}, {Braatz}, {Reid}, {Condon}, {Greene},
  {Henkel}, {Impellizzeri}, {Lo}, {Kuo}, {Pesce}, {Wagner}, \&
  {Zhao}}]{gao2017}
{Gao}, F., {Braatz}, J.~A., {Reid}, M.~J., {et~al.} 2017, \apj, 834, 52

\bibitem[{{Greene} {et~al.}(2010){Greene}, {Peng}, {Kim}, {Kuo}, {Braatz},
  {Impellizzeri}, {Condon}, {Lo}, {Henkel}, \& {Reid}}]{greene2010}
{Greene}, J.~E., {Peng}, C.~Y., {Kim}, M., {et~al.} 2010, \apj, 721, 26

\bibitem[{{Greene} {et~al.}(2013){Greene}, {Seth}, {den Brok}, {Braatz},
  {Henkel}, {Sun}, {Peng}, {Kuo}, {Impellizzeri}, \& {Lo}}]{greene2013}
{Greene}, J.~E., {Seth}, A., {den Brok}, M., {et~al.} 2013, \apj, 771, 121

\bibitem[{{Greene} {et~al.}(2016){Greene}, {Seth}, {Kim}, {L{\"a}sker},
  {Goulding}, {Gao}, {Braatz}, {Henkel}, {Condon}, {Lo}, \&
  {Zhao}}]{greene2016}
{Greene}, J.~E., {Seth}, A., {Kim}, M., {et~al.} 2016, \apjl, 826, L32

\bibitem[{{Greenhill} {et~al.}(1995){Greenhill}, {Jiang}, {Moran}, {Reid},
  {Lo}, \& {Claussen}}]{greenhill1995}
{Greenhill}, L.~J., {Jiang}, D.~R., {Moran}, J.~M., {et~al.} 1995, \apj, 440,
  619

\bibitem[{{G{\"u}ltekin} {et~al.}(2009){G{\"u}ltekin}, {Cackett}, {Miller}, {Di
  Matteo}, {Markoff}, \& {Richstone}}]{gueltekin2009}
{G{\"u}ltekin}, K., {Cackett}, E.~M., {Miller}, J.~M., {et~al.} 2009, \apj,
  706, 404

\bibitem[{{Heckman} {et~al.}(2005){Heckman}, {Ptak}, {Hornschemeier}, \&
  {Kauffmann}}]{heckman2005}
{Heckman}, T.~M., {Ptak}, A., {Hornschemeier}, A., \& {Kauffmann}, G. 2005,
  \apj, 634, 161

\bibitem[{{Herrnstein} {et~al.}(1999){Herrnstein}, {Moran}, {Greenhill},
  {Diamond}, {Inoue}, {Nakai}, {Miyoshi}, {Henkel}, \&
  {Riess}}]{herrnstein1999}
{Herrnstein}, J.~R., {Moran}, J.~M., {Greenhill}, L.~J., {et~al.} 1999, \nat,
  400, 539

\bibitem[{{Herrnstein} {et~al.}(1997){Herrnstein}, {Moran}, {Greenhill},
  {Diamond}, {Miyoshi}, {Nakai}, \& {Inoue}}]{herrnstein1997}
{Herrnstein}, J.~R., {Moran}, J.~M., {Greenhill}, L.~J., {et~al.} 1997, \apjl,
  475, L17

\bibitem[{{Kamali} {et~al.}(2017){Kamali}, {Henkel}, {Brunthaler},
  {Impellizzeri}, {Menten}, {Braatz}, {Greene}, {Reid}, {Condon}, {Lo}, {Kuo},
  {Litzinger}, \& {Kadler}}]{kamali2017}
{Kamali}, F., {Henkel}, C., {Brunthaler}, A., {et~al.} 2017, \aap, 605, A84

\bibitem[{{K{\"o}rding} {et~al.}(2006){K{\"o}rding}, {Falcke}, \&
  {Corbel}}]{koerding2006}
{K{\"o}rding}, E., {Falcke}, H., \& {Corbel}, S. 2006, \aap, 456, 439

\bibitem[{{Kuo} {et~al.}(2011){Kuo}, {Braatz}, {Condon}, {Impellizzeri}, {Lo},
  {Zaw}, {Schenker}, {Henkel}, {Reid}, \& {Greene}}]{kuo2011}
{Kuo}, C.~Y., {Braatz}, J.~A., {Condon}, J.~J., {et~al.} 2011, \apj, 727, 20

\bibitem[{{Kuo} {et~al.}(2013){Kuo}, {Braatz}, {Reid}, {Lo}, {Condon},
  {Impellizzeri}, \& {Henkel}}]{kuo2013}
{Kuo}, C.~Y., {Braatz}, J.~A., {Reid}, M.~J., {et~al.} 2013, \apj, 767, 155

\bibitem[{{Kuo} {et~al.}(2018){Kuo}, {Constantin}, {Braatz}, {Chung},
  {Witherspoon}, {Pesce}, {Impellizzeri}, {Gao}, {Hao}, {Woo}, \&
  {Zaw}}]{kuo2018}
{Kuo}, C.~Y., {Constantin}, A., {Braatz}, J.~A., {et~al.} 2018, \apj, 860, 169

\bibitem[{{Lal} {et~al.}(2004){Lal}, {Shastri}, \& {Gabuzda}}]{lal2004}
{Lal}, D.~V., {Shastri}, P., \& {Gabuzda}, D.~C. 2004, \aap, 425, 99

\bibitem[{{L{\"a}sker} {et~al.}(2016){L{\"a}sker}, {Greene}, {Seth}, {van de
  Ven}, {Braatz}, {Henkel}, \& {Lo}}]{lasker2016}
{L{\"a}sker}, R., {Greene}, J.~E., {Seth}, A., {et~al.} 2016, \apj, 825, 3

\bibitem[{{Masini} {et~al.}(2016){Masini}, {Comastri}, {Balokovi{\'c}}, {Zaw},
  {Puccetti}, {Ballantyne}, {Bauer}, {Boggs}, {Brandt}, {Brightman},
  {Christensen}, {Craig}, {Gandhi}, {Hailey}, {Harrison}, {Koss}, {Madejski},
  {Ricci}, {Rivers}, {Stern}, \& {Zhang}}]{masini2016}
{Masini}, A., {Comastri}, A., {Balokovi{\'c}}, M., {et~al.} 2016, \aap, 589,
  A59

\bibitem[{{Masini} {et~al.}(2017){Masini}, {Comastri}, {Puccetti},
  {Balokovi{\'c}}, {Gandhi}, {Guainazzi}, {Bauer}, {Boggs}, {Boorman},
  {Brightman}, {Christensen}, {Craig}, {Farrah}, {Hailey}, {Harrison}, {Koss},
  {LaMassa}, {Ricci}, {Stern}, {Walton}, \& {Zhang}}]{masini2017}
{Masini}, A., {Comastri}, A., {Puccetti}, S., {et~al.} 2017, \aap, 597, A100

\bibitem[{{Merloni} {et~al.}(2003){Merloni}, {Heinz}, \& {di
  Matteo}}]{merloni2003}
{Merloni}, A., {Heinz}, S., \& {di Matteo}, T. 2003, \mnras, 345, 1057

\bibitem[{{Middelberg} {et~al.}(2004){Middelberg}, {Roy}, {Nagar}, {Krichbaum},
  {Norris}, {Wilson}, {Falcke}, {Colbert}, {Witzel}, \&
  {Fricke}}]{middelberg2004}
{Middelberg}, E., {Roy}, A.~L., {Nagar}, N.~M., {et~al.} 2004, \aap, 417, 925

\bibitem[{{Miyoshi} {et~al.}(1995){Miyoshi}, {Moran}, {Herrnstein},
  {Greenhill}, {Nakai}, {Diamond}, \& {Inoue}}]{miyoshi1995}
{Miyoshi}, M., {Moran}, J., {Herrnstein}, J., {et~al.} 1995, \nat, 373, 127

\bibitem[{{Nagar} \& {Wilson}(1999)}]{nagar1999}
{Nagar}, N.~M. \& {Wilson}, A.~S. 1999, \apj, 516, 97

\bibitem[{{Nakai} {et~al.}(1993){Nakai}, {Inoue}, \& {Miyoshi}}]{nakai1993}
{Nakai}, N., {Inoue}, M., \& {Miyoshi}, M. 1993, \nat, 361, 45

\bibitem[{{Neufeld} \& {Maloney}(1995)}]{neufeld1995}
{Neufeld}, D.~A. \& {Maloney}, P.~R. 1995, \apjl, 447, L17

\bibitem[{{Neufeld} {et~al.}(1994){Neufeld}, {Maloney}, \&
  {Conger}}]{neufeld1994}
{Neufeld}, D.~A., {Maloney}, P.~R., \& {Conger}, S. 1994, \apjl, 436, L.127

\bibitem[{{Nisbet} \& {Best}(2016)}]{nisbet2016}
{Nisbet}, D.~M. \& {Best}, P.~N. 2016, \mnras, 455, 2551

\bibitem[{{Pesce} {et~al.}(2015){Pesce}, {Braatz}, {Condon}, {Gao}, {Henkel},
  {Litzinger}, {Lo}, \& {Reid}}]{pesce2015}
{Pesce}, D.~W., {Braatz}, J.~A., {Condon}, J.~J., {et~al.} 2015, \apj, 810, 65

\bibitem[{{Pjanka} {et~al.}(2017){Pjanka}, {Greene}, {Seth}, {Braatz},
  {Henkel}, {Lo}, \& {L{\"a}sker}}]{pjanka2017}
{Pjanka}, P., {Greene}, J.~E., {Seth}, A.~C., {et~al.} 2017, \apj, 844, 165

\bibitem[{{Pringle} {et~al.}(1999){Pringle}, {Antonucci}, {Clarke}, {Kinney},
  {Schmitt}, \& {Ulvestad}}]{pringle1999}
{Pringle}, J.~E., {Antonucci}, R.~R.~J., {Clarke}, C.~J., {et~al.} 1999, \apjl,
  526, L9

\bibitem[{{Rawlings} \& {Saunders}(1991)}]{rawlings91}
{Rawlings}, S. \& {Saunders}, R. 1991, \nat, 349, 138

\bibitem[{{Reid} {et~al.}(2009){Reid}, {Braatz}, {Condon}, {Greenhill},
  {Henkel}, \& {Lo}}]{reid2009}
{Reid}, M.~J., {Braatz}, J.~A., {Condon}, J.~J., {et~al.} 2009, \apj, 695, 287

\bibitem[{{Reid} {et~al.}(2013){Reid}, {Braatz}, {Condon}, {Lo}, {Kuo},
  {Impellizzeri}, \& {Henkel}}]{reid2013}
{Reid}, M.~J., {Braatz}, J.~A., {Condon}, J.~J., {et~al.} 2013, \apj, 767, 154

\bibitem[{{Saikia} {et~al.}(2015){Saikia}, {K{\"o}rding}, \&
  {Falcke}}]{saikia2015}
{Saikia}, P., {K{\"o}rding}, E., \& {Falcke}, H. 2015, \mnras, 450, 2317

\bibitem[{{Schmitt} {et~al.}(2002){Schmitt}, {Pringle}, {Clarke}, \&
  {Kinney}}]{schmitt2002}
{Schmitt}, H.~R., {Pringle}, J.~E., {Clarke}, C.~J., \& {Kinney}, A.~L. 2002,
  \apj, 575, 150

\bibitem[{{Sun} {et~al.}(2013){Sun}, {Greene}, {Impellizzeri}, {Kuo}, {Braatz},
  \& {Tuttle}}]{sun2013}
{Sun}, A.-L., {Greene}, J.~E., {Impellizzeri}, C.~M.~V., {et~al.} 2013, \apj,
  778, 47

\bibitem[{{Terlouw} \& {Vogelaar}(2015)}]{KapteynPackage}
{Terlouw}, J.~P. \& {Vogelaar}, M.~G.~R. 2015, {Kapteyn Package, version 2.3},
  {Kapteyn Astronomical Institute}, Groningen, available from
  \url{http://www.astro.rug.nl/software/kapteyn/}

\bibitem[{{White} {et~al.}(1997){White}, {Becker}, {Helfand}, \&
  {Gregg}}]{white1997}
{White}, R.~L., {Becker}, R.~H., {Helfand}, D.~J., \& {Gregg}, M.~D. 1997,
  \apj, 475, 479

\bibitem[{{Zhang} {et~al.}(2012){Zhang}, {Henkel}, {Guo}, \&
  {Wang}}]{zhang2012}
{Zhang}, J.~S., {Henkel}, C., {Guo}, Q., \& {Wang}, J. 2012, \aap, 538, A152

\bibitem[{{Zhao} {et~al.}(2018){Zhao}, {Braatz}, {Condon}, {Lo}, {Reid},
  {Henkel}, {Pesce}, {Greene}, {Gao}, {Kuo}, \& {Impellizzeri}}]{zhao2018}
{Zhao}, W., {Braatz}, J.~A., {Condon}, J.~J., {et~al.} 2018, \apj, 854, 124

\end{thebibliography}

\begin{appendix}
\section{The probability distribution of the observed PA offsets}\label{appendix:stat}
Detectable maser disks are edge-on so their normal lies in the plane of the sky. For the null hypothesis that jets have an isotropic distribution of orientations
in space (that is, they don't care about the maser disks), what is the probability distribution $\rm \rho(\delta)$ of the observed (projected onto the sky) PA offsets $\delta$ where 0 <$\delta$< $\pi$/2?
Let $\phi$ be the magnitude of the polar angle between the jet and the normal to the disk. Let $\theta$ be the azimuth angle in the disk plane, where $\theta$=0 corresponds to the line from the disk center pointing 
away from the observer. Thus $\delta$=$\phi$ $\rm sin(\theta)$. The differential probability of observing $\delta$ for a given $\phi$ is
\begin{equation}
 p_{\delta}(\delta|\phi)\,d\delta=p_{\delta}(\delta|\phi)\,\phi\, d sin(\theta)=p_{\theta}(\theta)\,d\theta=\frac{2\,d\theta}{\pi}
\end{equation}
so 
\begin{equation}
\begin{split}
 p_{\delta}(\delta|\phi)\,=\,\frac{2\,d\theta}{\pi\,\phi\,dsin\theta}\,=\,\frac{2}{\pi\,\phi\,cos\theta}\,=\,\frac{2}{\pi\,\phi\,(1-sin^2\theta)^{1/2}}\,=\\\,\frac{2}{\pi(\phi^2-\delta^2)^{1/2}}\, \, \quad  (\delta\,\leq\,\phi)
\end{split}
\end{equation}

For an isotropic distribution of jets, the probability distribution of polar angles $\phi$ is
$p_{\phi}(\phi)d\phi\,=\,sin\,\phi\,d\phi$.

The differential probability distribution of $\delta$ for all jets is obtained by integrating over $\phi$.
\begin{equation}
 p(\delta)=\int_{\delta}^{\pi/2}\, p_{\delta}(\delta|\phi) \, p_{\phi}(\phi)d\phi.
\end{equation}

Thus the differential distribution is 

\begin{equation}
 p(\delta)=\frac{2}{\pi}\int_{\delta}^{\pi/2}\, p_{\delta}\,\frac{sin\,\phi\,d\phi}{(\phi^2-\delta^2)^{1/2}}
\end{equation}

and the corresponding cumulative distribution is

\begin{equation}
 P(<\delta)\,=\,\int_{0}^{\delta}\, p(\acute{\delta})\,d\acute{\delta}.
\end{equation}

\begin{figure}[h!]
 \includegraphics[width=0.36\textwidth]{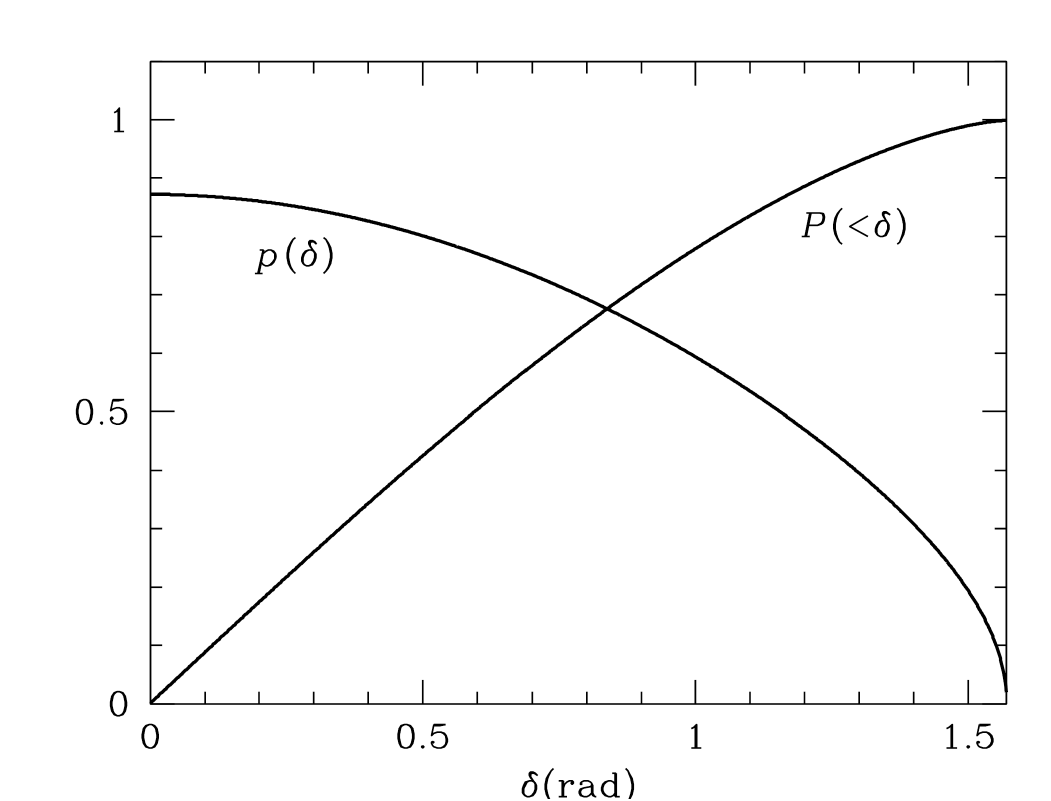}
 \caption{Differential probability distribution $p(\delta)$ of the PA difference $\delta$ between the disk normal and isotropic radio jets for edge-on maser disks and the cumulative distribution 
 $P(<\delta)$ of PA differences smaller than $\delta$.}\label{fig:probabilities}
\end{figure}

Both $p(\delta)$ and $P(<\delta)$ are plotted in Fig.\,\ref{fig:probabilities}.

\end{appendix}

\end{document}